\documentclass[journal]{IEEEtran}
\usepackage{amsmath}
\usepackage{amssymb}
\usepackage{color,soul}
\usepackage[pdftex]{graphicx}
\allowdisplaybreaks[4]
\usepackage[noadjust]{cite}
\usepackage{algorithm}
\usepackage{algorithmic}
\usepackage[caption=false, font=footnotesize]{subfig}

\usepackage{float}
\usepackage{color,soul}
\usepackage{eqnarray}

\usepackage{comment}

\newtheorem{example}{Example}

\usepackage{comment}
\usepackage[nolist]{acronym}

\newcommand{\q}{\mathsf{q}}
\renewcommand{\d}{\mathsf{d}}
\newcommand{\dq}{\mathsf{dq}}
\newcommand{\dqz}{\mathsf{dq0}}

\usepackage{pgfplots}
  \pgfplotsset{compat=newest}
  \usetikzlibrary{plotmarks}
  \usetikzlibrary{arrows.meta}
  \usepgfplotslibrary{patchplots}
  \usepackage{grffile}
  \usepackage{url}
  \usepackage{multirow}
  \usepackage{multicol}
  \usepackage{flushend}

\begin{document}

\title{Revisiting Power Systems Time-domain \\ Simulation Methods and Models}

\author{Jose Daniel Lara,~\IEEEmembership{Member,~IEEE}, Rodrigo Henriquez-Auba,~\IEEEmembership{Member,~IEEE},\\
Deepak Ramasubramanian,~\IEEEmembership{Senior Member,~IEEE}, Sairaj Dhople,~\IEEEmembership{Senior Member,~IEEE}, \\ Duncan S. Callaway,~\IEEEmembership{Member,~IEEE} and Seth Sanders,~\IEEEmembership{Fellow Member,~IEEE}
\thanks{Corresponding author: jdlara@berkeley.edu}
\thanks{This research was supported  by the U.S. Department of Energy by the Advanced Grid Modeling program within the Office of Electricity, under contract DE-AC02-05CH1123, and by the Solar Energy Technologies Office through award 38637 (UNIFI Consortium). The views expressed herein do not necessarily represent the views of the U.S. Department of Energy or the United States Government.}
\thanks{J. D.~Lara and D. S.~Callaway are with the Energy and Resources Group
at the University of California Berkeley.}
\thanks{R.~Henriquez-Auba, D. S.~Callaway and S. Sanders are with the Department of Electrical Engineering and Computer Sciences at the University of California Berkeley.}
\thanks{J. D.~Lara and R.~Henriquez-Auba are also with the National Renewable Energy Laboratory, Golden, Colorado.}
\thanks{D.~Ramasubramanian is with the Electric Power Research Institute, Knoxville, Tennessee.}
\thanks{S.~Dhople is with the Department of Electrical Engineering at University of Minnesota.}
\thanks{Manuscript received April 19, 2005; revised August 26, 2015.}}

\markboth{IEEE Transactions on Power Systems}%
{Lara \MakeLowercase{\textit{et al.}}: Revisiting the Assumptions of Power Systems Time-domain Simulation Methods}
%

\maketitle

\begin{abstract}
   The changing nature of power systems dynamics is challenging present practices related to modeling and study of system-level dynamic behavior. While developing new techniques and models to handle the new modeling requirements, it is also critical to review some of the terminology used to describe existing simulation approaches and the embedded assumptions. {This paper provides a first-principles review of the \textit{simplifications} and \textit{transformations} commonly used in the formulation of time-domain simulation models. It introduces a taxonomy and classification of time-domain simulation models depending on their frequency bandwidth, network representation, and software availability. Furthermore, it focuses on the fundamental aspects of averaging techniques, and model reduction approaches that result in modeling choices, and discusses the associated challenges and opportunities of applying these methods in systems with large shares of Inverter Based Resources (IBRs). The paper concludes with an illustrative simulation that compares the trajectories of an IBR-dominated system.}
\end{abstract}


\IEEEpeerreviewmaketitle
\begin{acronym}
\acro{EMT}{Electro-magnetic Transient}
\acro{DAE}{Differential Algebraic Equation}
\acro{IBR}{Inverter-based Resource}
\acro{ISO}{Indepent System Operator}
\acro{RMS}{Root Mean Square}
\acro{RCRF}{Rate of Change of the Reference Frame}
\acro{ODE}{Ordinary Differential Equation}
\acro{DE}{Differential Equation}
\acro{PLL}{Phase-locked Loop}
\acro{SPT}{Singular Perturbation Theory}
\acro{SW}{Switching Model}
\acro{AVG}{Average Value Model}
\acro{OEM}{Original Equipment Manufacturer}
\acro{SFA}{Shifted Frequency Analysis}
\acro{QSP}{Quasi-Static Phasor}
\acro{PDE}{Partial Differential Equation}
\acro{MOR}{Model Order Reduction}
\acro{VSM}{Virtual Synchronous Machine}
\end{acronym}
\section{Introduction}

\IEEEPARstart{T}{he} increasing integration of generation sources via power electronics is changing power systems. It is generally agreed that new dynamics in the controls of \acp{IBR}  change modeling requirements for system-wide stability studies that rely on time-domain simulations~\cite{paolone2020fundamentals, hatziargyriou2020definition, milano2018foundations}. In power systems dominated by synchronous generators, physical phenomena (magnetic fluxes, electro-mechanics, mechanical control reaction times, or thermo-dynamic processes) drive dynamic behavior. The control logic associated with synchronous generators is commonly tuned to the timescale of the relevant processes, creating a \emph{natural} separation between the dynamic behaviors attributable to physics and controls. On the other hand, \ac{IBR} dynamics are dominated by their controls, including modulation, \acp{PLL}, voltage, current, and power controllers. Therefore, the practical requirements of control design---often cascading PID control---defines the relationships between timescales. In fact, interactions at higher frequencies have recently been recognized with a new stability category~\cite{hatziargyriou2020definition}, highlighting the exigent need to revisit our understanding of system dynamics with high penetrations of \acp{IBR}. {The main goal of a time-domain simulation is to determine if there is a bounded trajectory of the system towards an equilibrium point following a disturbance. Determining the level of modeling detail required to capture the particular phenomena of interest accurately is key when developing a power systems simulation model.}

\begin{figure}[t]
    \centering
    \includegraphics[width=0.49\textwidth]{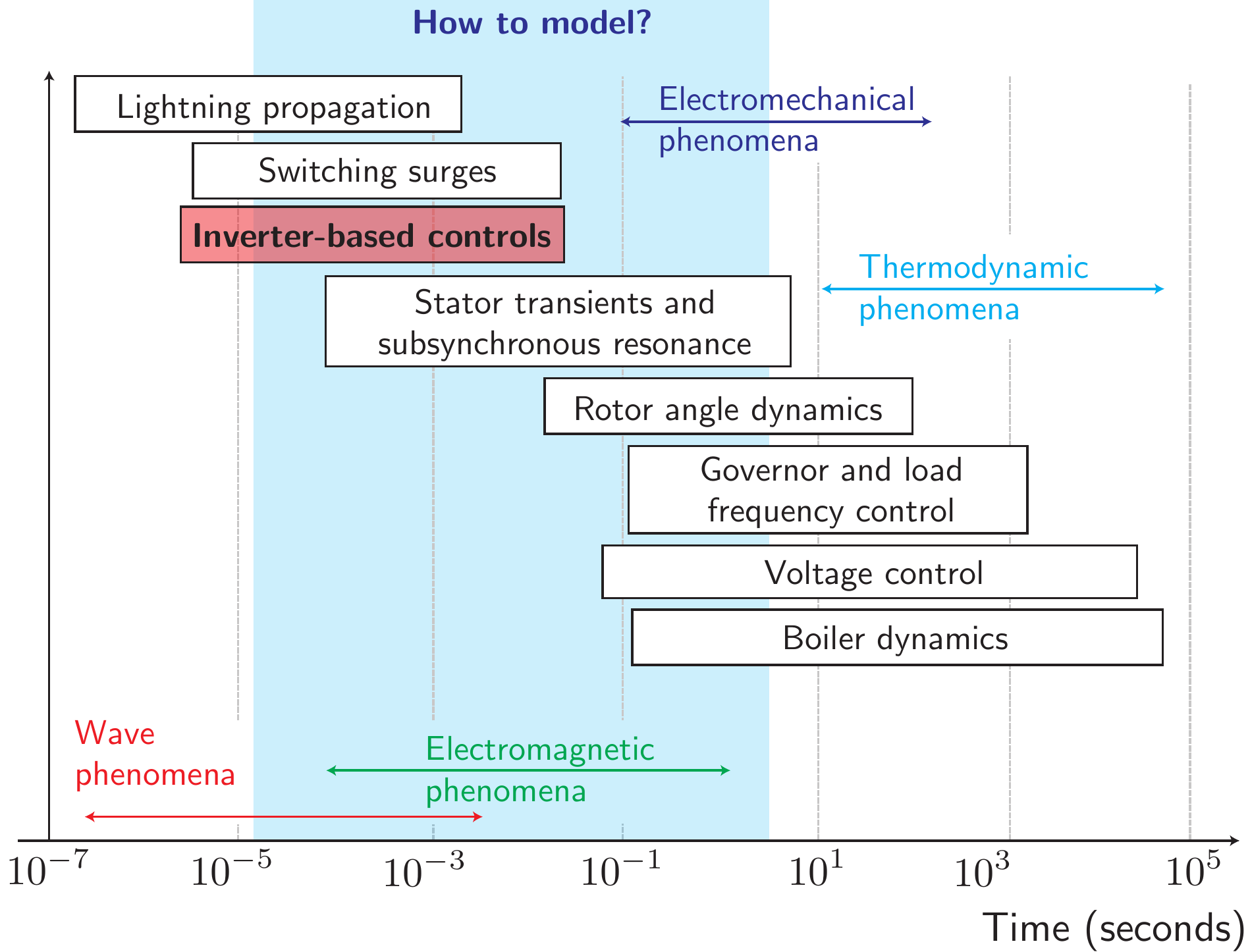}
\caption{{Time scales for power systems dynamic behavior. Synchronous machine-dominated systems exhibit dynamics dominated by physical laws. \acp{IBR} introduces new dynamics from logical components and pose new challenges in simulation model complexity for time-domain simulations~\cite{hatziargyriou2020definition}.}}
\label{fig:timescales}
\end{figure}

{Figure \ref{fig:timescales} shows the classification of power systems dynamic behaviors, as presented in~\cite{hatziargyriou2020definition}, and highlights the timescales over which \acp{IBR} interact with both electromechanical and electromagnetic phenomena.} Further, the modeling choices determine the algorithmic and computational requirements to execute the time-domain simulation {since analytical approaches do not typically scale for system-level studies}. In systems supplied predominantly by synchronous generators, IEEE Std 1110-2019~\cite{IEEE1110} guides the generator model complexity requirements based on the stability category under study and the severity of the perturbation. However, {up to now,} no such modeling guidance exists for \ac{IBR} dominated systems.

Several earlier papers have explored the modeling requirements for systems with \acp{IBR} and shed light on the theoretical~\cite{markovic2021understanding, misyris2021grid, henriquez2020grid} and practical~\cite{huang2020effect} requirements for time-domain simulation of large systems, particularly concerning the modeling of network circuit dynamics. Decisions about simulation models are often framed in terms of ``slow'' and ``fast'' dynamics, though formal definitions of slow and fast will depend on the context. This paper presents two formalizations of ``slow'' and ``fast'': in terms of a signal's bandwidth and in terms of model order reduction through \ac{SPT}.
The challenges of time-domain simulation in the context of \ac{IBR} dominated systems have been discussed recently. In \cite{paolone2020fundamentals}, the authors study the applicability of specific techniques to the simulation and modeling of systems with \acp{IBR}, from the signal processing perspective. A recent review ~\cite{9459192} discusses existing and novel methods to accelerate \ac{EMT} simulations in systems with \acp{IBR}. Composite system models for the network circuits, generators, and inverters of different types are described in~\cite{Green-2022a,venkatramanan2022integrated}.
%

\subsubsection*{Scope \& Contributions}
This paper seeks to clarify the semantics used to describe time-domain simulation models and their applications to system-wide simulations in the presence of \acp{IBR} given the need to develop novel simulation techniques that can capture \ac{IBR} dynamics more accurately. {While several efforts have focused on \acp{IBR} device-level simulation models \ \hbox{\cite{rosso2021grid, Sano2021simulation, lacerda2022inverter}}, this paper focuses on system-wide simulation models with an increased presence of \hbox{\acp{IBR}} and examines the implications for stability analysis. The objective is not to provide specific formulation or simulation-domain guidance but rather to interpret the assumptions and modeling approaches implicitly embedded in widely used simulation techniques and tools.} Particular emphasis is provided to reviewing \textit{Simplifications} and \textit{Transformations} inherent in various time-domain simulation methods and discussing the capabilities and limitations of these approaches. The contributions of the paper are as follows:
\begin{itemize}
    \item Provide a review of power systems time-domain simulation modeling approaches with applications to \acp{IBR}.
    \item Present a systematic discussion of the origins and underlying assumptions of different power systems dynamic modeling techniques, including application examples.
    \item Survey modeling approaches and terminology used in the literature and the level maturity of their implementation in commercial and research applications.
\end{itemize}

\begin{figure*}[t]
    \centering
    \includegraphics[width=0.93\textwidth]{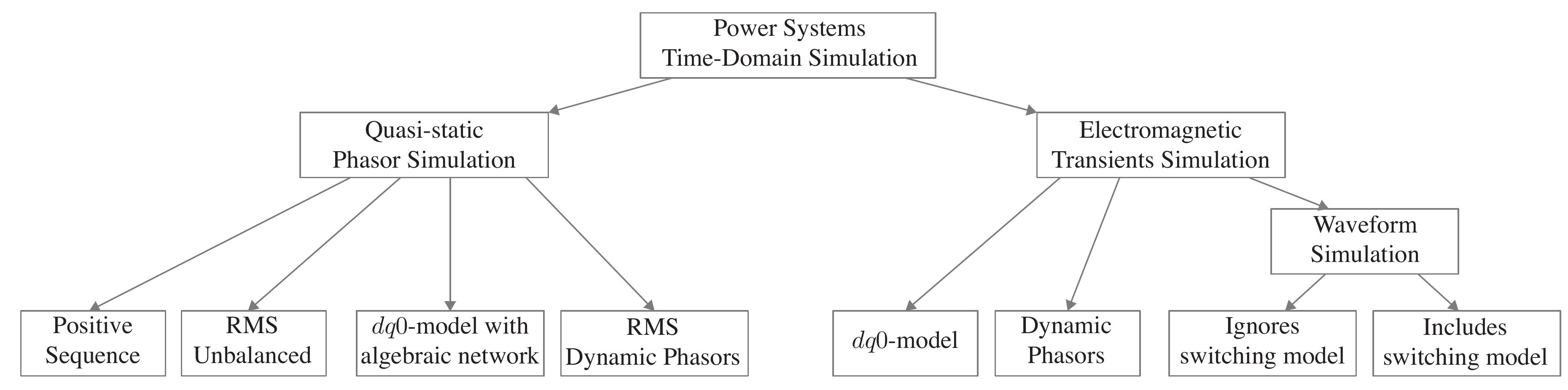}
    \caption{Taxonomy of power systems time-domain simulation models.}
    \label{fig:classification}
\end{figure*}

The paper is structured as follows: Section~\ref{sec:Definitions} discusses the definition of power systems time-domain simulation methods and introduces the taxonomy of the simulation models covered in this paper to facilitate the overall discussion. Section~\ref{sec:theory} reviews common simplifications and transformation techniques used in time-domain simulations relevant to defining the scope of simulation model definitions. Section~\ref{sec:Devices} discusses relevant aspects of network and device modeling in the context of time-domain simulation methods with examples of how simplifications and transformations are used. In Section~\ref{sec:names}, we discuss systems simulation focusing on definitions and categorizations, {and present an illustrative numerical test case}. Finally, conclusions are presented in Section~\ref{sec:Conclusions}.

\textbf{Notation:} Lower-case letters $x$ denote one-dimensional real variables, parameters, and functions; upper-case letters in the form $F$ and $F()$ are used for matrices and functions respectively; arrows (as in $\vec{x}$) represent vectors of variables or parameters; $\langle \cdot \rangle$ denote phasors; and lower-case letters in the form $\boldsymbol{x}$ denote complex-valued variables, functions, or parameters. Finally, $\jmath = \sqrt{-1}$.

\section{Time-domain Simulation} \label{sec:Definitions}

A time-domain simulation requires the specification of two ``layers'': (1)~a system model, including differential equations that describe the system's physics and controls, and (2)~a time stepping (or integration) algorithm. The choices made around the system model inform the integration algorithm requirements. This section defines what we mean by ``simulation model'' and provides a taxonomy of widely used methods.

\subsection{Definition}
A system simulation model is a set of dynamic equations in causal form for a collection of interconnected components expressed with explicit differential equations:
\begin{subequations}
\begin{align}
    \frac{d\vec{\boldsymbol{x}}(t)}{dt} &= F(\vec{\boldsymbol{x}}(t), \vec{\boldsymbol{y}}(t), \vec{\eta}, t), \quad \vec{\boldsymbol{x}}(t_0) = \vec{\boldsymbol{x}}^0, \label{eq:sim1}\\
    \frac{d\vec{\boldsymbol{y}}(t)}{dt} &= G(\vec{\boldsymbol{x}}(t), \vec{\boldsymbol{y}}(t), \vec{\psi}, t), \quad \vec{\boldsymbol{y}}(t_0) = \vec{\boldsymbol{y}}^0, \label{eq:sim2}
\end{align}
\end{subequations}
\noindent where $\vec{\boldsymbol{x}}(t)$ and $F(\cdot)$ represent the devices (e.g., \acp{IBR}, machines, loads) states and equations with $\vec{\eta}$ parameters. The circuit dynamics of the network are represented as the sub-system  $\vec{\boldsymbol{y}}(t)$ and $G(\cdot)$ with network parameters $\vec{\psi}$.\footnote{Simulation models could include a spatial component, as in the case of circuit models that consider traveling waves. In cases where the behavior of traveling waves is relevant, time-delay equivalents such as the Bergeron model are commonly used and can be implemented via \eqref{eq:sim1}-\eqref{eq:sim2}.} The simulation model \eqref{eq:sim1}-\eqref{eq:sim2} can be used to represent real- and complex-valued signal analysis.

Given the system model \eqref{eq:sim1}--\eqref{eq:sim2}, a simulation can be defined as follows: given an initial condition for the device and network states $\vec{\boldsymbol{x}}(t_0), \vec{\boldsymbol{y}}(t_0)$, advance the solution in time $t$ from one point to the next considering a discrete timeline $\{t_0,t_1,\dots,t_n,\dots,T\}$. A simulation requires a stepping algorithm that finds the solution at time $t_{n+1}$ given values of the involved variables over $\{t_0,t_1,\dots,t_n\}$. 

In steady state, voltage and current state variables in the model defined by~\eqref{eq:sim1}-\eqref{eq:sim2} are sinusoidally time-varying, making the dynamic response dependent on the value of $t$. Without some form of transformation, differential equations describing time-invariant system dynamics do not have well-defined equilibrium points. 
Solving time-varying nonlinear systems is expensive because integration algorithms use fixed time steps, and equilibrium points cannot be defined. Additionally, modeling three-wire three-phase power system components can result in intractable expressions with cross-coupling terms (for example, in the case of asymmetrical circuits or unbalanced signals).\footnote{The terminology adopted follows~\cite{yazdani2010voltage}, which uses \textit{balanced} in the context of signals and \textit{symmetric} in the context of circuits.} As we will see, \emph{simplifications} and \emph{transformations} influence the maximum required discretization step, $\Delta t$, and the number of states required to model the dynamics of interest.

\subsection{Taxonomy} \label{sec:taxonomy}

In power systems, time-domain simulation is roughly classified into two groups: 1) \textit{\ac{QSP} Simulations} in which the transmission system circuits dynamics are represented algebraically as discrete changes between steady-state operating points, and 2) \textit{\ac{EMT} Simulations} which include sufficient detail to capture fast dynamic phenomena. \ac{QSP}s are used for the study of low-frequency phenomena that ranges from inertial response to frequency regulation. On the other hand, \ac{EMT} simulations are used in settings where one wishes to capture the impact of line dynamics, converter switching, machine fluxes, and/or lightning surges.

Figure~\ref{fig:classification} depicts a taxonomy of time-domain simulation models found in the literature and covered in this paper. Though we have organized model types as subcategories of \ac{QSP} models and \ac{EMT} models, one can see there is a variety of different types of models within each category. We have organized model types roughly by the fastest time scale phenomena they are intended to capture. As we will explain in this paper, though modeling faster time scale phenomena requires more modeling detail, there are other important differences between these models.  Their mathematical formulations, the numerical algorithms available to numerically integrate them, and the assumptions required to interface machines and other energy-conversion interfaces across a network differ in important ways within the taxonomy.  These differences can introduce fundamental changes in model properties and output, influence the complexity of initializing model runs, and dictate the maximum allowable time step (and therefore, the computing time and data generated).

For practitioners, the definition of a simulation model is intertwined with the software environment used. Each simulation category (\ac{QSP}s or \ac{EMT}) has highly specialized models, algorithms, and modeling practices that have been developed over many decades. As a result, choosing a simulation model and its methods is inextricably tied to to the software environment and its capabilities. Therefore, in order to clearly talk about the difference between models in the taxonomy, we will introduce two broad classes of modeling choices: \emph{simplifications} and \emph{transformations}. The following section systematically reviews the broad space of potential simplifications and transformations. Subsequently, we discuss specific simulation tools in the context of these simplifications and transformations. This enables us to begin to understand the entire range of assumptions that go into different simulation tools and to move beyond the simple distinction of whether or not they are designed to study electromagnetic transients.

\section{Simplifications and Transformations}  \label{sec:theory}

By ``simplification'' and ``transformations,'' we mean mathematical manipulations of state variables that preserve the validity of some portion of a model's characterization of system physics and control loops. Transformations can have multiple equivalent formulations, and several works have developed a comprehensive analysis of transformations~\cite{milano2020frequency,o2019geometric} and arrived at equivalent conclusions to the ones presented here. Different from transformation, simplifications are aimed to reducing model complexity at the cost of an approximation to the real system dynamics.


\subsection{Simplification: Averaging Dynamics}
\label{sec:averaging}

Averaging methods focus on determining how the behavior of a complicated time-varying system can be approximated by a time-invariant system \cite{khalil2002nonlinear}. Averaging techniques are critical to reducing the computational cost of conducting simulations by reducing model bandwidth requirements, which, in turn, enables simulations to increase the maximum allowable time step $\Delta t$. {It is important to note that several quantities from a physical system can be averaged in a simulation model. For instance, averaging the commutation dynamics of the power electronics modules that operate in tenths of kHz bandwidth using a static representation \cite{yazdani2010voltage}. To this end, methods like Average-Value Modeling formalize the formulations that replace discontinuous switching with continuous models to accurately represent the converter's average behavior within switching intervals. In \cite{chiniforoosh2010definitions}, the authors provide a detailed review of converter switching averaging.
This paper focuses on the averaging of slower dynamics with bandwidth well below the kHz range that still challenge system level simulations.} Averaging fast dynamics not related to switching, the integration step of a power system model can go from 50ms to as large as one second for low-frequency dynamics \cite{zhang2007shifted}. Phasor representations of signals are regularly used as simplifications in power system simulations, and we discuss several approaches in the remainder of this subsection, along with their connection to averaging.

\subsubsection{Steady State Phasors} \label{sec:ss_phasors}
Steady state phasors are commonly defined in engineering as it applies to signals $s(t) = s_{pp} \cos(\varrho t + \theta) = \sqrt{2}~ \text{Re}\{S e^{\jmath (\varrho t +\theta)}\}$, on which the ``phasor'' $S = \frac{s_{pp}}{\sqrt{2}} e^{\jmath \theta}$ is expressed using the wave's root-mean-square value instead of the instantaneous value. This definition requires stationary conditions and homogeneous frequency\footnote{Refer to Section 1.4.1 \cite{milano2020frequency} and its references for further details on the origins of the term ``phasor.''}

\subsubsection{Dynamic Phasors} \label{sec:dynamic_phasor}
\textit{Dynamic phasor} as a concept was introduced in the early 1990s, has received renewed attention as a mechanism for simulating systems including fast \ac{IBR} dynamics. There are at least three independent but similar definitions of ``dynamic phasor'' in the literature, each developed using different approaches. The three approaches are motivated by the observation from averaging theory that the study of the signal envelope variations is sufficient to derive systems' stability properties.
Sanders \textit{et al.} \cite{sanders1991generalized} introduced \textit{dynamic phasors} via generalized averaging involving a time-dependent sliding-window interval $\mathcal{T}(t) = [t - T, t]$ for time-varying systems. In particular,\footnote{In this section $\varrho$ represents signal frequency, $\omega_s$ denotes system frequency.} consider the following expansion for a \textit{nearly periodic} signal $s(t)$:
\begin{align}
    s(t) = \sum_{k=-\infty}^\infty \langle s \rangle_k(t) ~ e^{\jmath k\varrho t}. \label{eq:dyn_phasor}
    \end{align}
The time-varying Fourier coefficients $\langle s \rangle_k(t)$, one for each harmonic component $k$, are recoverable as
    \begin{align}
    \langle s \rangle_k(t) = \frac{1}{T}\int_{\tau\in\mathcal{T}(t)} s(\tau) e^{-\jmath k\varrho\tau} \mathrm d\tau,
    \label{eq:dp_def}
\end{align}
\noindent and these are regarded as \textit{dynamic phasors}. The time dependency of $\langle s \rangle_k(t)$ enables the formulation of a differential rule to capture continuous time behavior as follows:
\begin{align}
    \left\langle\frac{d s(t)}{dt}\right\rangle_k =  \frac{d\langle s \rangle_k(t)}{dt} + \jmath k \varrho \langle s \rangle_k(t).
    \label{eq:dyn_phasor_diff}
\end{align}
\noindent extending dynamic phasors to complex periodic signals $s(t)$ is straightforward. For each $k\varrho$ where $\langle s \rangle_k(t) \ne \bar{0}$ there is a component at the negative frequency $-k\varrho$ with a phasor $\langle s \rangle_{-k}(t) = \langle s(t)^* \rangle^*_k$.
\noindent We will refer to this approach to dynamic phasors as the \textit{Fourier approach}. {A variation of the \textit{Fourier approach} described in this subsection is later employed in \cite{sudhoff1996transient} to analyze machine fed load-commutated converters.}

Another definition of dynamic phasors is derived from a communications-theoretic perspective~\cite{demarco_phasor_dynamics, venkatasubramanian1994tools, venkatasubramanian1995fast}. Venkatasubramanian introduced \textit{time-varying phasors} in~\cite{venkatasubramanian1994tools} defining such a phasor via a linear operator $P()$, such that $P(s(t)) = \langle s \rangle(t)$. In other words, $P()$ maps a signal $s(t)$ a phasor. The definition of \textit{time-varying phasors} begins with the idea that given a dynamic phasor of the form:
\begin{equation}
    \langle s \rangle(t) = s_{pp}(t) e^{\jmath \theta(t)},
    \label{eq:venka_dynp}
\end{equation}
\noindent there exist a unique band-pass modulated signal
\begin{equation}
    s(t) = \text{Re} \left\{ \langle s \rangle(t) e^{\jmath \varrho t} \right\} = s_{pp}(t)\cos(\varrho t + \theta(t)),
    \label{eq:signal}
\end{equation}
\noindent where $s_{pp}$ is the signal's amplitude. The operator $P()$ is defined for signals where it is possible to find a unique phasor $\langle s \rangle(t)$ such that $P^{-1} \left( P( s (t) ) \right) = s(t)$. The key property of $P()$ is that if the spectral content of $s(t)$ is slower than the carrier frequency $\varrho$, the operator maps to a unique phasor. Hence, the method is akin to a modulation operation to go from a time representation to a complex phasor representation.

From~\eqref{eq:signal}, it is possible to define the differential property of $P()$ as follows:
\begin{equation}
    P \left( \dfrac{ds(t)}{dt} \right )  = \dfrac{d\langle s \rangle(t)}{dt} + \jmath \varrho \langle s \rangle(t).
    \label{eq:ventka_diff}
\end{equation}
\noindent The differential property in \eqref{eq:ventka_diff} is the same as the Fourier dynamic phasor in \eqref{eq:dyn_phasor_diff} for $k = 1$. We will refer to this approach to defining a dynamic phasor as the \textit{linear operator approach}.

The third approach to deriving dynamic phasors is rooted in the \ac{SFA} concept developed by Martí \textit{et al.} \cite{marti2005shifted, zhang2007shifted}, and it is closely related to the definition of the linear operator $P()$ introduced in \cite{venkatasubramanian1994tools}.\footnote{In fact, Venkatasubramanian mentions that the time-varying phasor approach is inspired by the Hilbert transform, which is central to SFA.} \ac{SFA} begins with the assumption that a power system dynamic signal is a band-pass, real-valued, and can be represented by a Fourier decomposition as follows:
\begin{align}
    s(t) &= s_{pp,d}(t)\cos(\varrho t) - s_{pp,q}(t)\sin(\varrho t),
    \label{eq:sfa_signal}
\end{align}
\noindent and the dynamic phasor is defined as $\langle s \rangle(t) = s_{pp,d}(t) + \jmath s_{pp,q}(t)$. It is composed of low-pass functions in quadrature $s_{pp,d}(t) = s_{pp}(t)\cos(\theta(t))$ and $s_{pp,q}(t) = s_{pp}(t)\sin(\theta(t))$. The procedure described by \ac{SFA} provides a method to obtain $\langle s \rangle(t)$ through a \textit{shifted frequency} representation of $s(t)$:
\begin{equation}
    \langle s \rangle(t) = \left ( s(t) + \jmath \text{H} \left \{ s(t) \right \} \right )e^{-\jmath \varrho t},
    \label{eq:sfa_1}
\end{equation}
\noindent where $\text{H}()$ is the Hilbert transform~\cite{haykin2008communication}. Equation \eqref{eq:sfa_1} shifts signal $s(t)$ with the form \eqref{eq:sfa_signal} by $-\varrho$. As a result, it centers all the signal's dynamics around $0$ Hz. The differential property for \eqref{eq:sfa_1} can be derived from differentiating \eqref{eq:sfa_1} resulting in \eqref{eq:ventka_diff}.
In each case, the definition of ``phasor'' corresponds to the complex envelope of a signal with a modulating frequency $k\varrho$ (with $k=1$ only in the linear operator and \ac{SFA} approaches, and arbitrary integer $k$ in the Fourier approach). The three approaches reach the same operation for the differentiation of phasorial representation of a signal and its properties.

The definitions above provide several properties required to leverage dynamic phasors for time-domain simulation of average models. First, note that a dynamic phasor operation is interpreted as a band-pass modulation \cite{venkatasubramanian1994tools, zhang2007shifted} where a given time-varying signal $s(t)$ is demodulated into a set of dynamic phasors $\langle s \rangle_k(t)$ with a given carrier frequency $\varrho$. The implication is that the signal $s(t)$ needs to possess a well-defined Fourier transform, which requires that: 1) The integral of $s(t)$ needs to be well-defined over the period $\mathcal{T} = [t - T, t]$, i.e.,~\eqref{eq:dp_def} needs to bounded, and 2) for a given frequency $k \varrho$, the signal $s(t)$ is narrow-band with a bandwidth $\mathcal{B} = [(k-1)\varrho, (k+1) \varrho]$~\cite{demiray2008simulation}. This demodulation perspective provides a definition of ``slow'' and ``fast'' dynamics of the system with respect to the modulating frequency $\varrho$ and its harmonics as noted in \cite{demarco_phasor_dynamics}. In particular, the bandwidth (or speed) of the dynamic phasor is limited by the modulating signal frequency $k \varrho$ in order for the reverse transformation to be one-to-one. This discussion highlights that under certain conditions, it is possible to use an average model of the dynamics without any information loss.

Using dynamic phasors for simulation model averaging comes with potential trade-offs since the increase in the number of states can dilute the gains derived from using larger $\Delta t$~\cite{demiray2008simulation}. This is particularly problematic in cases where the resulting average models are still time-varying. The practical implication of these limitations is that the dynamic phasor representation of a high-frequency (i.e., fast) signal could require many $k$ components. It also implies that the representation of high-frequency dynamic behavior using a single carrier frequency may introduce error in the demodulation process~\cite{zaborszky1993error}.

\subsubsection{Representation of three-phase signals and models}

The definitions in Section~\ref{sec:dynamic_phasor} focus on ``single-phase'' complex signals. However, there is no unique application of dynamic phasors to poly-phase systems, contributing to the diversity of definitions and formulations when the technique is used for time-domain simulation. For instance, \ac{SFA} and the linear operator are applied phase-by-phase to a three-phase $abc$ in~\cite{zhang2007shifted, hart2019impact, venkatasubramanian1994tools} to study unbalanced systems. In this case, each phase needs to comply with the requirements described in Section~\ref{sec:dynamic_phasor} to enable an exact representation of the underlying signals. Another implementation is used in~\cite{demiray2008simulation}, where the Fourier approach is used to develop models for each $k$\textsuperscript{th} harmonic required when modeling machines, FACTS, and HVDC-links. Similarly, the Fourier dynamic phasors were implemented in conjunction with sequence transformations in~\cite{986446} to study unbalanced AC machines. The myriad implementations of dynamic phasors to three-phase signals and systems make it difficult to point out a universally accepted set of properties and limitations.

Transmission-level power systems analysis is generally concerned with modeling multi-phase system quantities with a common \emph{system frequency} $\omega_s$ and an offset $\theta$ for each phase. For simplicity in the discussion that follows, we consider a three-wire three-phase signal $\boldsymbol{s}_{abc}(t)$ (that satisfies $\boldsymbol{s}_{abc}(t)^\top \mathbf{1} = 0$) of the form:
\begin{align}
    \vec{s}_{abc}(t) = \begin{bmatrix} s_{a}(t) \cos(\omega_s t + \theta_a(t)) \\
    s_{b}(t) \cos(\omega_s t + \theta_b(t)) \\
    s_{c}(t) \cos(\omega_s t + \theta_c(t)) \end{bmatrix},
    \label{eq:three_phase}
\end{align}
\noindent where $s_{a}(t),$ $s_{b}(t),$ and $s_{c}(t)$ are real-valued wave amplitudes, $\theta_{a}(t),$ $\theta_{b}(t),$ and $\theta_{c}(t)$ represent the phase difference between the waves, and the signal's frequency is the \emph{system frequency} $\omega_s$, which is commonly assumed as a uniform ``system frequency". $\boldsymbol{s}_{abc}(t)$ can be used to represent voltages, currents, electromagnetic flux or physical values like inductance.

In this paper, we approach the study of dynamic phasors for three-phase signals using a \emph{space vector} representation that provides a complex signal equivalent derived as follows:
\begin{align} 
    \!\!\boldsymbol{s}(t) \!=\! c \!\! \begin{bmatrix} s_{a}(t)\!\!\left(e^{\jmath (\omega_s t + \theta_a(t))} \!\!+ \!\!e^{-\jmath (\omega_s t + \theta_a(t))} \right)\\
    s_{b}(t)\!\!\left(e^{\jmath (\omega_s t + \theta_b(t))} \!\!+ \!\!e^{-\jmath (\omega_s t + \theta_b(t))} \right)\\
    s_{c}(t)\!\!\left(e^{\jmath (\omega_s t + \theta_c
    (t))} \!\!+ \!\!e^{-\jmath (\omega_s t + \theta_c(t))} \right) \end{bmatrix}^\top \!\!\! \begin{bmatrix} e^{\jmath 0} \\
    e^{\jmath \frac{2\pi}{3}} \\
    e^{\jmath \frac{-2\pi}{3}} \end{bmatrix},
    \label{eq:three_phase_complex}
\end{align}
\noindent where the constant $c$ is set to scale the vector's magnitude. For an arbitrary three-phase signal,~\eqref{eq:three_phase_complex} permits representation in a two-dimensional vector as follows:~\cite{yazdani2010voltage,o2019geometric} {(see Chapter 4 in}~\cite{yazdani2010voltage} {and}~\cite{o2019geometric}):
\begin{equation}
    \boldsymbol{s}(t) = c \left(  \boldsymbol{s}_+(t)e^{\jmath \omega_s t}  +  \boldsymbol{s}_-(t)e^{-\jmath \omega_s t} \right)
    \label{eq:three_phase_complex_signal}
\end{equation}
\noindent where the terms $\boldsymbol{s}_+(t)$ and $\boldsymbol{s}_-(t)$,
\begin{align}
    \boldsymbol{s}_+(t) &= s_{a}(t) e^{\jmath \theta_a(t)} + s_{b}(t)e^{\jmath( \theta_b(t)+\frac{2\pi}{3})} \notag \\ &\quad + s_{c}(t)e^{\jmath(\theta_c(t) -\frac{2\pi}{3})} \label{eq:positive_phasor} \\
    \boldsymbol{s}_-(t) &= s_{a}(t) e^{-\jmath \theta_a(t)} + s_{b}(t)e^{-\jmath( \theta_b(t)-\frac{2\pi}{3})} \notag \\ &\quad+ s_{c}(t)e^{-\jmath (\theta_c(t) +\frac{2\pi}{3})}, \label{eq:negative_phasor}
\end{align}
\noindent are complex-valued multipliers for the positive and negative frequency components, respectively. This shows that any three-phase signal without a zero-sequence component and with frequency $\omega_s$ can be modeled without loss of information using two complex-value dynamic phasors as long as $\boldsymbol{s}_+(t),\boldsymbol{s}_-(t)$ comply with the requirements described in Section~\ref{sec:dynamic_phasor}. This approach is used in~\cite{986446} where the authors develop a model for signals of the form~\eqref{eq:three_phase_complex_signal} for an unbalanced machine using~\eqref{eq:dyn_phasor}-\eqref{eq:dyn_phasor_diff} with $k=1$ and $k=-1$ components.

The implication relevant to time-domain simulation is that it is not possible to directly model an arbitrary three-phase signal~\eqref{eq:three_phase} with one complex phasor quantity--put formally: $\langle s(t) \rangle \ne \langle s(t) \rangle_{+} + \langle s(t) \rangle_{-}$. Nevertheless, if the signal is balanced--, i.e., $ s_{a}(t) =  s_{b}(t) =  s_{c}(t) = s(t)$ and $\theta_a(t) = \theta(t)$ $\theta_b(t) = \theta_a(t) - \frac{2\pi}{3}$ $\theta_c(t) = \theta_a(t) + \frac{2\pi}{3}$--then components $\boldsymbol{s}_+(t)$, $\boldsymbol{s}_-(t)$ reduce to:
\begin{align}
    \boldsymbol{s}_+(t) = 3 s(t) e^{\jmath \theta(t)}, \quad \boldsymbol{s}_-(t) = 0, \label{eq:balanced_phasor}
\end{align}
\noindent which implies that $\langle s(t) \rangle = \langle s(t) \rangle_{+}$. The result in~\eqref{eq:balanced_phasor} showcases the effectiveness of averaging techniques in balanced systems: \emph{the envelope of a balanced three-phase signal is completely determined by the positive-frequency phasor}. {Handling unbalanced three-phase signals with dynamic phasors usually implies layering transformations as presented in \cite{hart2019impact, demiray2008simulation} where the authors employ unbalanced-three-phase to sequence components via the Fortescue transform \cite{fortescue} and finally get the dynamic phasors for the positive and negative components.}
However, even if a dynamic simulation requires modeling phasors for components at multiple frequencies, it can improve the solution times with respect to modeling directly in $abc$ if the increase in $\Delta t$ compensates for the increase in the number of states.

\subsection{Reference Frame Transformations \label{sec:theory_transformations}}

{The application of reference frame transformations in simulation models is ubiquitous since they are the basis of rotating electric machinery analysis and power electronic converter control. However, for developing simulation models, Park's Transform (or $\dqz$ transform) \cite{park1929two} is the most commonly employed. Park's original use case was the transformation of the machine's rotor and the stator rotational mutual inductance to a time-invariant quantity. The applicability of the transform has been expanded to other devices and network quantities}. Park's transform is defined as:
\begin{align}
 \vec{s}_{dq0}(t) = C {T}_p(\theta(t)) \vec{s}_{abc}(t)
\label{eq:Park_transform}
\end{align}

\noindent where $\theta(t) $ is the angle between the ``reference axes'' and the axes of rotation. {The transform's matrix $T_p(\theta(t))$ can be defined arbitrarily for any $\theta(t)$, which leads to different formulations depending on the leading-lagging relationships between the $\d$ and $\q$ axis.}
This paper follows the convention in Standard IEEE-1110 \cite{IEEE1110}.
whereby the $\q$-axis lags the $\d$-axis and the $a$-phase is aligned with the $\d$-axis to define the transformation matrix:
\begin{equation}
\small{
    \!\!\!{T}_p(\theta(t)) \!=\!\! \begin{bmatrix} \cos\theta(t) & \cos(\theta(t) \!-\! \frac{2\pi}{3})  & \cos(\theta(t) \!+\! \frac{2\pi}{3}) \\ - \!\sin\theta(t) & -\!\sin( \theta(t) \!-\! \frac{2\pi}{3}) & -\! \sin(\theta(t) \!+\! \frac{2\pi}{3}) \\ 1 & 1  & 1  \end{bmatrix}
    }.
    \label{eq:Park_matrix}
\end{equation}
While manipulations involving dynamic phasors hinge on assumptions concerning the frequency content involved in signals, the transform~\eqref{eq:Park_transform} can be applied to any three-phase signal (in both directions). Although the reference transform is valid and reversible for any combination of $\boldsymbol{s}_{abc}(t)$ and $\theta(t)$, transforming the system is \textit{useful} for the purposes of simulation only with careful selection of $\theta(t)$. {As discussed in \cite{o2019geometric}, the choice for $\theta(t)$ can yield transformation matrices equal to the ones used in other transformations. For example, with $\theta(t) = \theta_0$, the $\alpha\beta 0$/Clarke transformation is obtained. These transformations are widely applied in control design for converters or torque control for induction motors but are less useful for simulation model formulation applications.}

In simulation models, the calculation of the angle $\theta(t)$ is usually done by integrating the frequency of the signal \eqref{eq:three_phase}:
\begin{equation}
    \theta(t) = \int_{t_o}^t \left(\omega_s + \Delta\omega_s(\tau)\right) \mathrm d\tau + \theta^0.
    \label{eq:park_angle}
\end{equation}
\noindent where, respectively, $\omega_s$ and $\Delta\omega_s(t)$ are the frequency and its time-dependent variation. In a balanced system, the effectiveness of the transformation relies on using a reference frame with a constant frequency, i.e., $\Delta\omega_s(t) = 0$, such that we can obtain:
\begin{align}
 \vec{s}_{dq0}(t) = c {T}_p(\theta(t)) \vec{s}_{abc}(t) \triangleq \boldsymbol{s}_+(t)
\label{eq:Park_dynamic_phasor}
\end{align}
\noindent Under balanced conditions, Park's transform maps the three-phase signal's dynamic phasor $\boldsymbol{s}_+(t)$. The transformation reduces the simulation models by transforming the variables of the components to a rotating reference frame~\cite{belikov2017comparison} which requires fewer variables without loss of information.

In interconnected device simulations, we can define the reference frame to reference transformation:
\begin{align}
    {T}_p(\theta_1(t)) {T}_p(\theta_2(t))^{-1} \!=\!\! \begin{bmatrix} \cos \Delta \theta(t) & \sin \Delta \theta(t)  & 0 \\
    \!-\! \sin \Delta \theta(t) & \cos \Delta \theta(t)  & 0 \\
    0 & 0  & 1  \end{bmatrix}
\end{align}
\noindent where $\Delta \theta(t) = \theta_1(t) - \theta_2(t) = \int_{t_0}^t \left(\Delta\omega_1(\tau) - \Delta\omega_2(\tau)\right) d\tau + \Delta \theta^0$. The above property implies that only the local reference frames'  frequency deviations are required for the system simulation. Moreover, the selection of common system frequency $\omega_s$ is of no consequence to the simulation results as long as there is a local $\Delta\omega(t)$ variable at the device model.

In the case of machine models, it is possible to define the change of the local reference frame frequency in terms of the rotor speed changes, i.e., $\Delta\omega(t) = \Delta \omega_r(t)$. This property was identified in \cite{zaborszky1993error}, where the authors recognized that the machine rotor is a \emph{perfect demodulator of the network signals}. This implies that each synchronous machine imposes, through its rotor angular speed, the frequency at the bus. Further, the ``system frequency'' is not a modeled quantity, nor is it required to represent the system dynamics accurately.

However, unlike electrical machine models, an \ac{IBR} does not necessarily have a local reference frame. Hence, when simulating \acp{IBR}, the relationship of the device model with $\omega_s$ is dictated by the controls and their representation. For instance, many \ac{IBR} control structures rely on frequency measurements at the local bus for the internal controls, introducing a heretofore unproblematic requirement in system simulations: namely, reliably modeling frequency dynamics at the bus. When modeling control schemes that require frequency measurements, the idealized frequency measurement is modeled as follows:
\begin{equation}
    \Delta \omega_\mathrm{bus}(t) = \frac{d}{dt}\tan^{-1} \frac{V_{q,\mathrm{bus}}(t)}{V_{d,\mathrm{bus}}(t)}.
    \label{eq:bus_frequ}
\end{equation}
\noindent The implication of having to estimate the local frequency in a simulation is that~\eqref{eq:bus_frequ} relies on the value of the voltage states to estimate the local bus frequency. As a result, the frequency deviations' impact on the simulation becomes dependent on the network model's assumptions. Further, modeling frequency measurement with~\eqref{eq:bus_frequ} suggests that simplifications in the network dynamics can mischaracterize the effects of frequency deviations on the controls and introduce further deviations from the real-world behavior of the \ac{IBR}.

Recent works have looked at the theoretical underpinnings of the system's modeling and the definition of system frequency in simulation models to develop novel theories about the relationships between energy and system frequency \cite{milano_complex, milano2020frequency} to develop a rigorous perspective that considers the inclusion of \acp{IBR}. However, this is still an area under active development, and there are not widely adopted practices to manage the frequency selection for large-scale simulations.

\subsection{Singular perturbation theory (SPT) \label{sec:SPT}}

In the power systems literature, it is common to make approximations that reduce the number of states modeled~\cite{stott1979power}. The justifications for several reductions are based on practical knowledge; however, these approximations have also been formalized in terms of time-scale separation arguments derived from \ac{SPT} of system dynamics~\cite{chow1982time, kokotovic1999singular}. \footnote{The reader is referred to~\cite{kokotovic1999singular} for the foundations of \ac{SPT}.}
In the context of the model~\eqref{eq:sim1}--\eqref{eq:sim2}, \ac{SPT} identifies two state categories: ``slow'' states ($\vec{\boldsymbol{x}}_s$,  $\vec{\boldsymbol{y}}_s$) and ``fast'' states ($\vec{\boldsymbol{x}}_f$, $\vec{\boldsymbol{y}}_f$). Fast states are multiplied by a small positive real scalar $\varepsilon$ that represents all the small parameters to be neglected. The dynamics are then presented as:
\begin{equationarray}{rclrcl}
    \dot{\vec{\boldsymbol{x}_s}} &\hspace{-0.2cm}=&\hspace{-0.2cm} F_s(\vec{\boldsymbol{x}}, \vec{\boldsymbol{y}}, \vec{\eta}, \varepsilon),& \quad \dot{\vec{\boldsymbol{y}_s}} &\hspace{-0.2cm}=&\hspace{-0.2cm} G_s(\vec{\boldsymbol{x}}, \vec{\boldsymbol{y}}, \vec{\psi},  \varepsilon)  \label{eq:spt_slow}\\
    \varepsilon\dot{\vec{\boldsymbol{x}_f}} &\hspace{-0.2cm}=&\hspace{-0.2cm} F_f(\vec{\boldsymbol{x}}, \vec{\boldsymbol{y}}, \vec{\eta}, \varepsilon),& \quad
    \varepsilon\dot{\vec{\boldsymbol{y}_f}} &\hspace{-0.2cm}=&\hspace{-0.2cm} G_f(\vec{\boldsymbol{x}}, \vec{\boldsymbol{y}}, \vec{\psi}, \varepsilon) \label{eq:spt_fast}
\end{equationarray}

\ac{SPT} involves finding the trajectories of a dynamical system as $\varepsilon \to 0$. If the conditions are met, it is possible to set $\varepsilon = 0$ and reduce~\eqref{eq:spt_fast} to algebraic equations. The requirement to perform a reduction in the number of dynamic states, is that the algebraic equations~\eqref{eq:spt_fast} in the domain of interest have distinct roots {(i.e., the model is in \textit{standard form} \cite{kokotovic1999singular})}:
\begin{align}
   \vec{\boldsymbol{x}}_f = R_{x}(\vec{\boldsymbol{x}}_s, \vec{\boldsymbol{y}}_s, \vec{\eta}),\quad \vec{\boldsymbol{y}}_f = R_{y}(\vec{\boldsymbol{x}}_s, \vec{\boldsymbol{y}}_s, \vec{\psi}). \label{eq:spt_root}
\end{align}
If the roots of \eqref{eq:spt_root} are distinct, it is possible to guarantee, via eigenvalue analysis, that during a transient the fast dynamics will not diverge. As described in~\cite{kokotovic1999singular}, most of this analysis is done on the ``boundary layer system'' that imposes requirements on the eigenvalues of the Jacobian of $F_f$ and $G_f$. {Distinct isolated roots enable the use of variable substitution which results in a well-defined reduced model at a specific root value as follows}:
\begin{subequations}
\begin{align}
    \dot{\vec{\boldsymbol{x}_s}} &= F_s(\vec{\boldsymbol{x}}_s, \vec{\boldsymbol{y}}_s, R_{x}(\vec{\boldsymbol{x}}_s, \vec{\boldsymbol{y}}_s, \vec{\eta}), \vec{\eta}), \label{eq:spt_device_transformed_f}\\
    \dot{\vec{\boldsymbol{y}_s}} &= G_s(\vec{\boldsymbol{x}}_s, \vec{\boldsymbol{y}}_s, R_{y}(\vec{\boldsymbol{x}}_s, \vec{\boldsymbol{y}}_s, \vec{\psi}), \vec{\psi}). \label{eq:spt_network_transformed_g}
\end{align}
\end{subequations}
System \eqref{eq:spt_device_transformed_f}-\eqref{eq:spt_network_transformed_g} is the \textit{quasi-steady-state model}. These definitions based on \ac{SPT} provide a different definition of ``slow'' and ``fast'' in a simulation, this time, in terms of the ``speed'' of some states with respect to others. \ac{SPT} provides conditions under which the fast states may rapidly converge to a root that approaches the final value of the state, i.e., $\vec{\boldsymbol{x}}_f \to R_{x}(\vec{\boldsymbol{x}}_s, \vec{\boldsymbol{y}}_s, \vec{\eta})$ and $\vec{\boldsymbol{y}}_f \to R_{y}(\vec{\boldsymbol{x}}_s, \vec{\boldsymbol{y}}_s, \vec{\psi})$.

The theoretical underpinnings of \ac{SPT} justify the practice of neglecting line capacitance and the resulting voltage dynamics in short-length lines due to the small value of the current drawn from the shunt impedance. \ac{SPT} is used in \cite{sauer1988integral, sauer2017power} to show that modeling current flows over the network via the admittance matrix $\boldsymbol{Y}$ is a reasonable approximation of the integral manifold. Such approximations have also been shown to be valid for small signal analysis and are typically used for studying intra- and inter-area electromechanical oscillations, as well as stabilizer parameter tuning \cite{sauer2017power}.





\section{Network and Device Modeling}  \label{sec:Devices}

The device and network models' level of detail determines the requirements placed on time-domain simulation solution methods. The simulation models must accurately capture the interactions across the different timescales. Thus, both network and device dynamic models formulations must be consistent in their assumptions.

\subsection{Network modeling} \label{sec:Network}

As discussed in Section \ref{sec:taxonomy}, the representation of network circuit dynamics in the simulation models is a major distinguishing attribute in establishing the taxonomy. 
The transmission line model formulation has significant importance on the time-domain simulation category selection since it impacts the bandwidth~\cite{belikov2017comparison} and time-constant of the states~\cite{henriquez2020grid} added to the model.

The detailed dynamic model of a transmission line is a model using~\acp{PDE} for the voltage and current across the line length with distributed parameters. However, it is difficult to include this level of detail in time-domain simulation tools, and experience shows that the impacts of using simplified models are not significant for most applications. The appropriateness of using simplifying assumptions depends on the line length: \emph{long} ($>200$ km), \emph{medium} ($80$-$200$ km), and \textit{short} ($<80$ km)~\cite{sauer2017power}. In \emph{short} lines, shunt elements are not considered ($C'=G'=0$), reducing the model to a series $r$-$l$ circuit~\cite{sauer2017power}. Another common simplification is the lossless line assumption ($R'=G'=0$), which results in a simplified two-port model interfaced via a current source with time delay $\tau = d\sqrt{L'/C'}$ (where $d$ is the line length)~\cite{dommel1969digital}. The Bergeron line model extends the delayed current-source model approximating the series resistance losses by adding lumped resistances at both ends of the two-port system.

The lumped parameter $\pi$-circuit of a transmission line results from a quasi-static analysis when voltage and currents are assumed to be sinusoidal functions with constant frequency. The use of \acp{ODE} of the $RLC$ $\pi$-model is a \emph{simplification} of the \acp{PDE}, and the accuracy of this approach depends on the line parameters and specific phenomena studied~\cite{sauer2017power}. The following example showcases the transformation for the $\pi$ line model and the application of simplifications.

\begin{example}
Consider the three-phase lumped-circuit elements $\pi$-line model in Fig.~\ref{fig:network_diagram}. The per-unit voltage and line dynamics of a symmetrical circuit are captured by \acp{ODE}:
\begin{subequations}
\begin{align}
    \frac{l}{\Omega_b} \frac{d}{dt}\vec{i}_{abc,\ell}(t) &= [\vec{v}_{abc,2}(t) - \vec{v}_{abc,1}(t)] - r\vec{i}_{abc,\ell}(t), \label{eq:current_abc} \\
    \frac{c}{\Omega_b} \frac{d}{dt}\vec{v}_{abc,1}(t) &= [\vec{i}_{abc,1}(t) - \vec{i}_{abc,\ell}(t)] - g\vec{v}_{abc,1}(t), \label{eq:voltage_abc} \\
    \frac{c}{\Omega_b} \frac{d}{dt}\vec{v}_{abc,2}(t) &= [\vec{i}_{abc,\ell}(t) - \vec{i}_{abc,2}(t)] - g\vec{v}_{abc,2}(t), \label{eq:voltage2_abc}
\end{align}
\label{eq:line_full_model}
\end{subequations}
in which $\Omega_b$ is the base frequency in rad/s. {The value of  $\Omega_b$ depends on the specific system under study, it can take the values $2 \pi 60$ or $2 \pi 50$ rad/s}.  $\vec{i}_{abc,1}(t)$ and $\vec{i}_{abc,2}(t)$ are three-phase real-valued quantities related to adjoining dynamical systems. The dynamic $\pi$-model model \eqref{eq:line_full_model} has nine states, and it is commonly used on waveform simulations (see Fig. \ref{fig:classification}) and as the base to develop transformed line models.

\begin{figure}[t]
    \centering
    \includegraphics[width=0.35\textwidth]{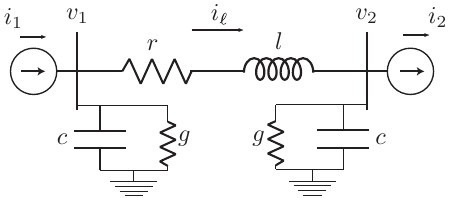}
    \caption{One-line diagram of a connection between two three-phase sources.}
    \label{fig:network_diagram}
\end{figure}

The system of equations~\eqref{eq:current_abc}-\eqref{eq:voltage2_abc} can be transformed using Park's transformation~\eqref{eq:Park_transform} with a fixed frequency $\omega_s$ to yield the following dynamics in the $\dq$ reference frame:
\begin{subequations}
\begin{align}
    \frac{l}{\Omega_b} \frac{d}{dt}\boldsymbol{i}_\ell(t) &=  \left[\boldsymbol{v}_{1}(t) - \boldsymbol{v}_{2}(t)\right] - (r + \jmath \omega_s l) \boldsymbol{i}_\ell(t),  \label{eq:dq_current} \\
    \frac{c}{\Omega_b}\frac{d}{dt}\boldsymbol{v}_{1}(t) &=  \left[\boldsymbol{i}_{1}(t) - \boldsymbol{i}_\ell(t)\right] - (g + \jmath \omega_s c) \boldsymbol{v}_{1}(t), \label{eq:dq_voltage}
    \\
    \frac{c}{\Omega_b}\frac{d}{dt}\boldsymbol{v}_{2}(t) &=  \left[\boldsymbol{i}_\ell(t) - \boldsymbol{i}_{2}(t)\right] - (g + \jmath \omega_s c) \boldsymbol{v}_{2}(t), \label{eq:dq_voltage2}
\end{align}
\end{subequations}
\noindent where $\boldsymbol{v}(t) = v_\d(t) + \jmath v_\q(t)$, $\boldsymbol{i}(t) = i_\d(t) + \jmath i_\q(t)$, $\boldsymbol{i}_{1}(t)$, $\boldsymbol{i}_{2}(t)$, and $\boldsymbol{i}_{\ell}(t)$ are the transformed $\vec{i}_{abc,1}(t)$, $\vec{i}_{abc,2}(t)$, $\vec{i}_{abc,\ell}(t)$ output currents. If the signals $\vec{i}_{abc,1}(t)$ and $\vec{i}_{abc,2}(t)$ are balanced, the resulting model eliminates any sinusoidal time-varying characteristic and reduces the model to six time-invariant differential equations. Under balanced conditions,~\eqref{eq:dq_current}--\eqref{eq:dq_voltage2} are, together, the dynamic phasor representation of the network quantities, reflecting the common approach to capturing line dynamics in time-domain simulations as presented in Fig.~\ref{fig:classification}. If $\vec{i}_{abc,1}(t)$ or $\vec{i}_{abc,2}(t)$ is unbalanced or contains harmonics, the $\dq$ transformation yields time-varying signals, and the benefits could be minimal with respect to $abc$ modeling as in~\eqref{eq:current_abc}-\eqref{eq:voltage2_abc} {The results of this transformation do not generalize to arbitrary unbalanced three-phase signals. For unbalanced signals, it is required to follow the approach in \cite{hart2019impact} where the authors first obtain balanced equivalents for each positive and negative sequence components and afterwards introduce the $\dqz$ transform.}

The transformed \ac{ODE} system \eqref{eq:dq_current}-\eqref{eq:dq_voltage2} can be further simplified employing \ac{SPT} since the terms $\ell/\Omega_b,c/\Omega_b \approx 10^{-3}$ have small values in most systems operating at 50 or 60 Hz. Making the left-hand-side of \eqref{eq:dq_current}-\eqref{eq:dq_voltage2} zero yields the following algebraic map:
\begin{align}
    \begin{bmatrix}
    \boldsymbol{i}_{1} \\
    -\boldsymbol{i}_{2} \end{bmatrix} = \begin{bmatrix}
    \boldsymbol{y}_\ell + \boldsymbol{y}_c & -\boldsymbol{y}_\ell \\ -\boldsymbol{y}_\ell & \boldsymbol{y}_\ell + \boldsymbol{y}_c \end{bmatrix} \begin{bmatrix} \boldsymbol{v}_{1} \\ \boldsymbol{v}_{2}
    \end{bmatrix}, \label{eq:network_approx}
\end{align}
where $\boldsymbol{y}_l = (r+\jmath \omega_s l)^{-1}$ and $\boldsymbol{y}_c = g + \jmath \omega_s c$. This model formulation is equivalent to commonly used {current-injection  representation of the network model} $G(\cdot)$ \cite{venkatasubramanian1994tools, venkatasubramanian1995fast, belikov2017comparison}:
\begin{subequations}
\begin{align}
    \frac{d\vec{\boldsymbol{x}}(t)}{dt} &= F(\vec{\boldsymbol{x}}(t), \vec{\boldsymbol{v}}(t)), \label{eq:sim3_phasor}\\
    0 &= I(\vec{\boldsymbol{x}}(t), \vec{\boldsymbol{v}}(t)) - \boldsymbol{Y} \vec{\boldsymbol{v}}(t). \label{eq:sim4_phasor}
\end{align}
\end{subequations}
{If the devices' model $F(\vec{\boldsymbol{x}}(t), \vec{\boldsymbol{v}}(t))$ only accounts for low-frequency neglecting high-frequency dynamics like stator fluxes or fast controls, the use of simplifications and transformations turns the time-variant dynamic model~\eqref{eq:sim1}--\eqref{eq:sim2} to an index-1 \ac{DAE}\footnote{Index is a notion used in the theory of \acp{DAE} for measuring the distance from a \ac{DAE} to its equivalent representation as an \ac{ODE}. Index-1 \acp{DAE} can be solved in principle by substituting the values of the algebraic variable in the differential equations \cite{Petzold82dae}.}} Note that this formulation coincides with the single-phase steady-state phasor representation commonly used for {\ac{QSP} (or ``positive sequence'') simulations; we will provide details on this taxonomy in Section~\ref{sec:categories}}.
\end{example}

{The accuracy of the reduced order model depends on the term that multiplies the derivative ($\ell/\Omega_b$ and $c/\Omega_b$ in \eqref{eq:dq_current}-\eqref{eq:dq_voltage2}) and its proximity to zero. This approximation error is typically observed in the system's dynamic response over a disturbance near time zero \cite{sauer1988integral}.} {If $\varrho = \omega_s$, this representation yields a quasi-static model of the network (also known as \emph{time-varying phasor}) representation of the network \cite{venkatasubramanian1994tools, venkatasubramanian1995fast, baimel2017dynamic}.}

The model representation in~\eqref{eq:sim3_phasor}-\eqref{eq:sim4_phasor} has been justified in practice since it improves computational performance in large system transient stability studies without a significant loss of accuracy~\cite{zaborszky1993error, stott1979power}. The relative error introduced by~\eqref{eq:sim4_phasor} is proportional to the ratio of the variation with respect to the carrier frequency. For instance, rotor dynamics in the range of 2 Hz introduce an error of around 3\%. As a result, simulation tools use models that ignore the dynamics of the electrical network and should not be used to investigate phenomena such as sub-synchronous resonance, control interactions, and oscillations or phenomena with dynamics $>$ 10 Hz~\cite{zaborszky1993error}.
\vspace{-0.3cm}
\subsection{Synchronous Generator Modeling}

\begin{figure}[t]
    \centering
    \includegraphics[width=0.49\textwidth]{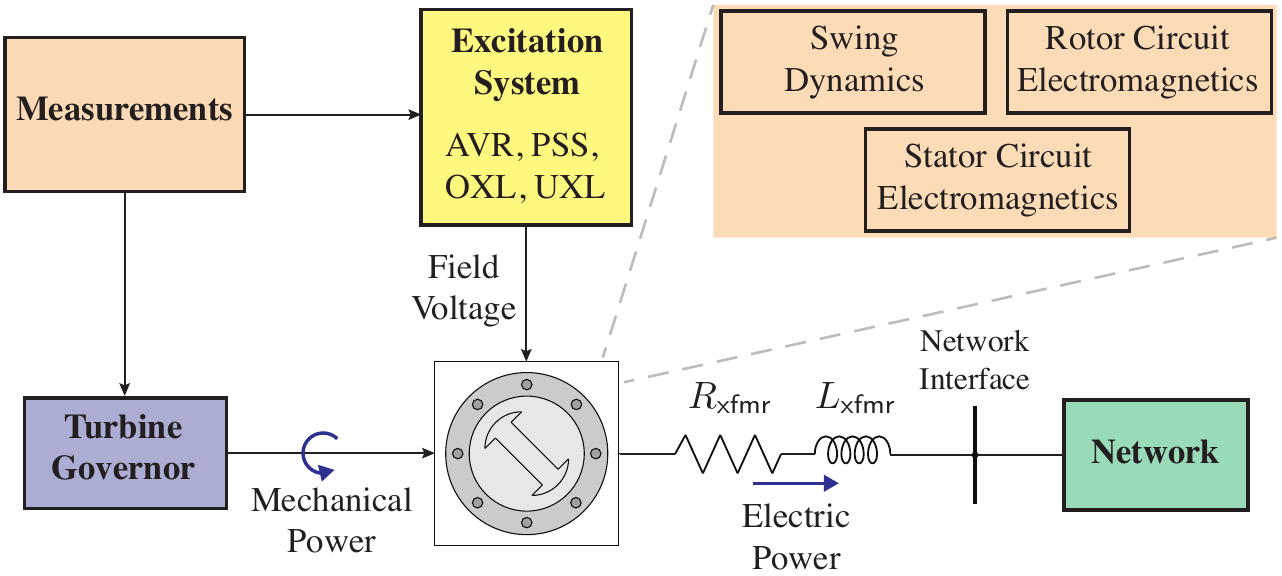}
    \caption{Simplified synchronous generator block diagram.}
    \label{fig:generator_diagram}
\end{figure}

A typical disaggregation of a simulation model for a synchronous generator depicting mechanical, electrical, and electromagnetic dynamics is shown in Fig.~\ref{fig:generator_diagram}. The $abc$ three-phase model of a synchronous generator includes time-varying inductance terms due to the rotation of the rotor's electromagnetic field. The application of Park's transform significantly reduces model complexity by eliminating time-varying terms~\cite{park1929two,sauer2017power}. The resulting machine model is composed of stator and rotor voltage equations, rotor flux linkage equations, and the rotor swing dynamics~\cite{sauer2017power}.

\begin{example} Although it is possible to perform simplifications in any portion of the generator model, machine models are the prime candidate for reducing the simulation model's complexity. Take the stator voltage equations in $\dq$ assuming a balanced system and $\omega_r \approx 1$:
\begin{align}
    \frac{1}{\Omega_b} \frac{d}{dt}{\boldsymbol{\psi}}_{\dq} &= \boldsymbol{e}_{\dq} + r_a \boldsymbol{i}_{\dq} - \jmath \boldsymbol{\psi}_{\dq}. \label{eq:stator_flux}
\end{align}
In a machine model, the base frequency term $\Omega_b^{-1} \approx 10^{-3}$ is relatively small in systems operating at 50 or 60 Hz. Hence, employing \ac{SPT}, it is possible to set the term $\Omega_b^{-1} \dot{\psi}_{\dq} = 0$ and reduce the flux dynamics~\eqref{eq:stator_flux}. The simplification implies that the terms $\dot{\psi}_{\dq}$ decay very rapidly after a perturbation, as observed in practice. This simplification also removes high-frequency transients and fundamental frequency components in the currents $i_\d$ and $i_\q$, which in turn also reduces the bandwidth of the model allowing larger time steps in the simulation execution. {Once high-frequency terms are eliminated it is possible to include other approximations to reduce model complexity. For instance, the simplifying assumption that the per-unit value of $\omega_r$ is a fixed value 1.0 in equations \eqref{eq:stator_flux}}.
This simplification can be taken one step further via variable substitution of $\psi_{dq}$ in the computation of the air-gap torque directly using voltages and currents as follows:
\begin{align}
    \tau_e &= \psi_d i_\q - \psi_\q i_\d = (e_\q + r_a i_\q)i_\q - (-e_\d - r_a i_\d)i_\d \notag \\
    &= e_\d i_\d + e_\q i_\q + r_a(i_\d^2 + i_\q^2) \label{eq:airgap_torque}
\end{align}
\end{example}
The resulting model has the property of representing the energy balance at the shaft of the machine in terms of a \ac{QSP} representation of the power at the stator.
\begin{figure}[t]
    \centering
    \includegraphics[width=0.49\textwidth]{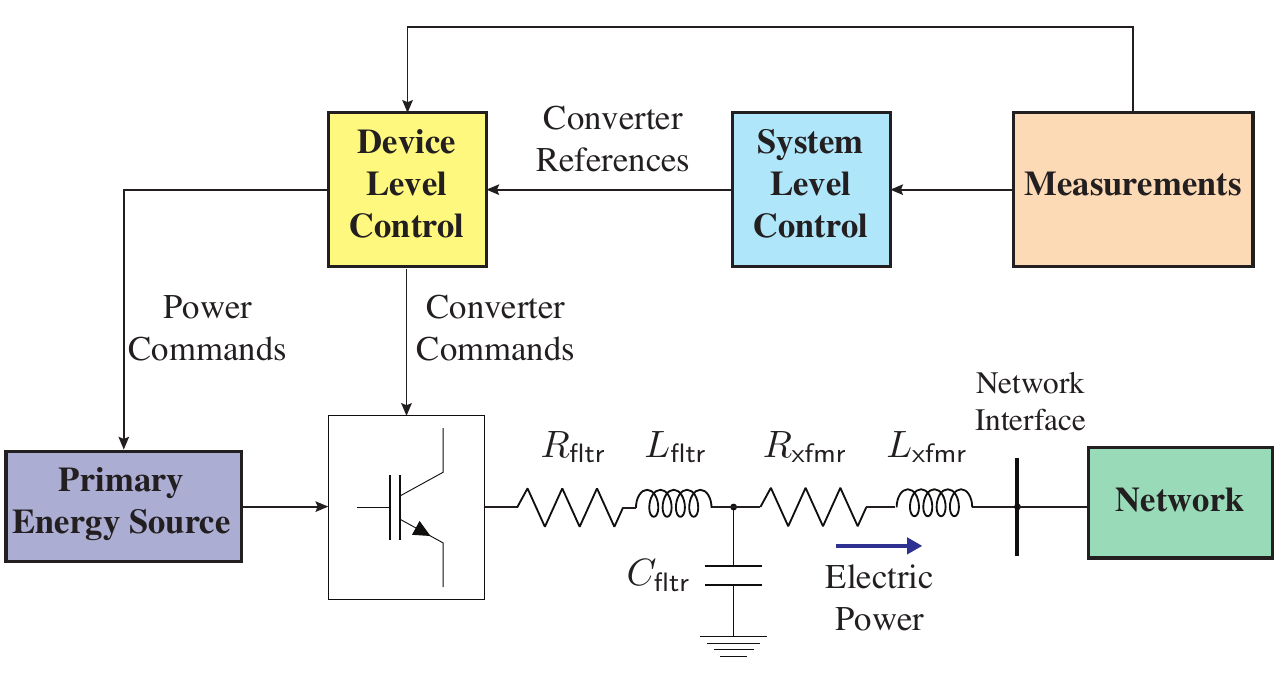}
    \caption{Simplified converter block diagram.}
    \label{fig:converter_diagram}
\end{figure}

\vspace{-0.3cm}
\subsection{Inverter Based Resources (IBR) Modeling}

Energy conversion is fundamentally different in \acp{IBR} compared to synchronous generators. Instead of using a rotational magnetic field, \acp{IBR} synthesizes voltages through high-frequency switching, introducing high-frequency dynamics into the synthesized voltages and realized currents. Despite the high-frequency behavior, \acp{IBR} include output filters and controllers with low-pass filters tuned to regulate average values; these aspects readily provide the rationale to disregard switching-level dynamics from \ac{IBR} models  intended for system-level studies~\cite{yazdani2010voltage}. 

\ac{IBR} simulation models are structured according to the cascaded control architectures used in the field (see Fig.~\ref{fig:converter_diagram}). \ac{IBR} controls employ transformations to reduce the number of control loops and produce and regulate three-phase signals (of the form~\eqref{eq:three_phase}); in this setting, transformations facilitate avoiding the regulation of sinusoidal signals, which requires high-order controls~\cite{yazdani2010voltage}. \acp{IBR} controls commonly employ Park's Transform since it has the property of reducing the controls' bandwidth requirements in order to achieve zero steady-state error. Hence, the model developer needs to be more aware of the effects that simplifications could have, not just on the model precision but also on the reliability of the control performance~\cite{venkatramanan2022integrated}.

Models used to represent \ac{IBR} dynamic behavior can either be \ac{OEM} specific or generic~\cite{pourbeik2017genericmodels,Ajala22HiCSS}. The wide variety of control architectures and topologies poses a challenge in the development of simulation models for \acp{IBR}. The level of detail used to model each sub-component depends on the study requirements, and--as in the case of generators--there needs to be consistency between the device timescales and the remainder of the system model. For instance, in \ac{EMT} \ac{IBR} models, the dynamic behavior of the filter is commonly represented in detail, given that, in \ac{EMT} simulation, the dynamics of the lines are also included. On the other hand, in \ac{QSP} models, \ac{IBR} ``fast'' controls may not be included if the rest of the system model's time constants are significantly larger than the controller's time constants. The fact that \acp{IBR} model dynamics can extend over multiple timescales depending on the control architecture and the time constants makes it challenging to determine whether a particular control can be reduced or should be modeled in detail. For this reason, the performance of a model needs to be verified against equipment behavior before integration into existing software tools. A contemporary example of such a transition is the \ac{IBR} model for low short circuit systems~\cite{ramasubramanian2020IBRlowshortckt} which has now been named as the REGC\_C model under the generic suite of models for \acp{IBR}. Note that the intent of this paper is not to indicate that \ac{OEM} or generic models belong to a specific category of simulation; both models can be implemented for quasi-static or \ac{EMT} simulations. Rather, the intent is to assert that device models need to match the overall system assumptions.

\begin{example}
 One of the possible simplifications for converter models in power system studies relates to local voltage and current controllers. As described in~\cite{luo2014spatiotemporal, markovic2021understanding}, the internal controller involves two cascading PI controllers:
 \begin{subequations}
 \begin{align}
      k_{iv}^{-1} \frac{d}{dt} \boldsymbol{\phi} &= - k_{pv}^{-1} \boldsymbol{\phi} +  k_{pv}^{-1}k_{iv}^{-1} (\boldsymbol{i}_\text{cv}^\text{ref} - \jmath C_\text{fltr} \boldsymbol{v}_o), \label{eq:spt_inverter1}\\
      k_{ic}^{-1} \frac{d}{dt} \boldsymbol{\gamma} &= -k_{pc}^{-1} \boldsymbol{\phi} +  k_{pc}^{-1}k_{ic}^{-1} (\boldsymbol{v}_\text{cv}^\text{ref} - \jmath L_\text{fltr} \boldsymbol{i}_\text{cv}),
      \label{eq:spt_inverter2}
 \end{align}
 \end{subequations}
\noindent where $\boldsymbol{\phi}$ are the integrator states (for $\d$ and $\q$ axes) of the voltage controller, intended to track the capacitor voltage $\boldsymbol{v}_o$ to a reference provided from the system level control $\boldsymbol{v}_o^\text{ref}$, while $\boldsymbol{\gamma}$ are the integrator states of the current controller, intended to track the converter current output $\boldsymbol{i}_\text{cv}$ to a reference provided from the voltage controller $\boldsymbol{i}_\text{cv}^\text{ref}$. The output of these controllers will output the converter voltage command $\boldsymbol{v}_\text{cv}^\text{ref}$. Similar to Example~1, simplifications of the device level controllers dynamics can be argued from \ac{SPT} since the integral gains $k_{iv}^{-1}, k_{ic}^{-1}$ can be significantly small to due to timescale separation requirements~\cite{luo2014spatiotemporal,Ajala-2021}. In practice, this simplification of inner controls is equivalent to assuming that system-level control references are tracked without error.

Eliminating the internal current controls can also be extended to the filter dynamics in certain settings~\cite{Ajala-2021}. Since the filters are $RLC$ circuits designed to attenuate high-frequency switching phenomena, they usually exhibit time constants around $10^{-3}$s and can be eliminated using the same procedure as in Section~\ref{sec:Network}. The resulting simplified model is a controlled voltage source behind a transformer impedance, where the voltage values are determined by the system-level control dynamics.
\end{example}

\section{Simulation Categories}  \label{sec:categories}

\begin{table*}[t]
\centering
\caption{Simulation category review and classification}
\label{tab:category}
\begin{tabular}{|c|c|c|c|c|}
\hline
\textbf{\begin{tabular}[c]{@{}c@{}}Simulation\\ Category\end{tabular}}                              & \textbf{Name}                                                                                            & \textbf{\begin{tabular}[c]{@{}c@{}}Multiple Frequencies /\\ Asymmetrical / Unbalanced\end{tabular}} & \textbf{\begin{tabular}[c]{@{}c@{}}Network\\ Representation\end{tabular}} & \textbf{\begin{tabular}[c]{@{}c@{}}Software\\ Availability\end{tabular}} \\ \hline \hline
\multirow{4}{*}{\begin{tabular}[c]{@{}c@{}} \\[3ex] Quasi-static\\  phasor Simulation\end{tabular}}           & \begin{tabular}[c]{@{}c@{}}\textbullet~ Positive Sequence \\ \textbullet~ RMS Balanced \\ \textbullet~Electromechanical$^\dagger$\\ \textbullet~Transient Stability$^\dagger$\end{tabular}                      & No                                                                                     & Single Phase                                                                  & Mature                                                              \\ \cline{2-5} 
                                                                                                    & \textbullet~ RMS Unbalanced                           & No$^*$                                                                                                  & Symmetrical Components                                                                   & Mature                                                                   \\
                                                                                                    \cline{2-5} 
                                                                                                    & \begin{tabular}[c]{@{}c@{}} \textbullet~$\dqz$-model with \\ ~~ algebraic network\end{tabular} & Yes                                                                                                 & $\dqz$                                                            & Research                                                                       \\                                                                                              \cline{2-5} 
                                                                                                    & \begin{tabular}[c]{@{}c@{}}\textbullet~RMS Dynamic Phasor\end{tabular} & Yes                                                                                                 & $abc$ or $\dqz$                                                              & Research                                                           \\ \hline
\multirow{3}{*}{\begin{tabular}[c]{@{}c@{}}Electromagnetic\\ Transients (EMT)\\ Simulation\end{tabular}} & \begin{tabular}[c]{@{}c@{}}\textbullet~Waveform\\ \textbullet~Point-on-Wave\\ \textbullet~$abc$-model\end{tabular}                     & Yes                                                                                                 & $abc$                                                                & Mature                                                                                                                 \\ \cline{2-5} 
                                                                                                    & \begin{tabular}[c]{@{}c@{}}\textbullet~$\dqz$-model\\ \textbullet~Fast Time-Varying Phasor\end{tabular}                                                                                             & Yes                                                                                                 & $\dqz$                                                                 & Research                                                                      \\ \cline{2-5} 
                                                                                                    & \textbullet~Dynamic Phasor                                                                                           & Yes                                                                                                 & $abc$ or $\dqz$                                                              & Research                                                              \\ \hline
\multicolumn{5}{p{15cm}}{$^\dagger$These are commonly used for balanced systems, but they can be formulated for unbalanced networks as RMS Unbalanced model.} \\
\multicolumn{5}{p{15cm}}{$^*$The unbalanced network is handled algebraically by connecting the symmetrical circuits into a single phase.} 
\end{tabular}
\end{table*}

Table~\ref{tab:category} summarizes common simulation categories and associated characteristics based on the taxonomy in Fig.~\ref{fig:classification}. The model properties are split between the larger categories of \ac{QSP} and Electromagnetic Transients (\ac{EMT}) given the commonalities in modeling assumptions. The main difference between the two categories stems from the representation of the network circuits. We have identified the modeling of the network circuits as the main determinant of a simulation model's numerical properties, available solution techniques and resulting computational requirements.
\label{sec:names}

\vspace{-0.3cm}

\subsection{\ac{QSP}}

\ac{QSP} models are narrow-band simulation models that focus on dynamics that do not deviate significantly from the steady-state frequency. In most software implementations, the integration method independently solves the device and network equations using an algebraic representation of~\eqref{eq:sim2}.

\subsubsection{Positive Sequence} Also called \textit{\ac{RMS} balanced} simulation; this is a \ac{QSP} simulation where the balanced network is modeled using a single phase, the positive sequence. These models are formulated in the form of \acp{DAE} of the general form \eqref{eq:sim3_phasor}-\eqref{eq:sim4_phasor} and use solution strategies derived from the Partitioned-Explicit solution methods, relying on the assumption that the differential portion of the \ac{DAE} is not stiff. These types of simulators are primarily designed for the analysis of machine angles in balanced transmission systems. This simulation category is also commonly called \textit{Electromechanical} or \textit{Transient Stability} simulations.
\subsubsection{RMS Unbalanced} This is a simulation used to analyze non-symmetrical network circuits. The network is modeled using symmetrical components while keeping the representation algebraic. In this simulation, the dynamic components feed into the positive sequence subset of equations, and the three sequence networks are simplified using circuit algebra depending on the type of fault~\cite{venkatasubramanian1994tools}. These models are commonly used to evaluate the system's transient behavior during and after unbalanced faults.
\subsubsection{\ac{RMS} Dynamic Phasors} This approach is based on constructing the model using Fourier Dynamic Phasor as defined in Section \ref{sec:dynamic_phasor}. {This approach employs the same right-hand-side from \eqref{eq:dq_current}-\eqref{eq:dq_voltage2} to model the network circuit. However, the derivative terms (left-hand side) are neglected since, in most RMS models, they represent faster dynamics than the devices' models.} This model produces a subset of equations \eqref{eq:sim3_phasor}-\eqref{eq:sim4_phasor} for each $k$\textsuperscript{th} harmonic component that can be integrated as a system of \acp{DAE}. The model's accuracy depends on the number of harmonics considered in the model and whether the dynamics around each harmonic possess the required properties to neglect the left-hand side derivative \cite{demiray2008simulation}.

\subsubsection{$\dqz$-model with algebraic network} This type of simulation model is obtained by performing a $\dqz$ transformation and assuming a steady state (e.g., by setting the left-hand side of~\eqref{eq:dq_current}--\eqref{eq:dq_voltage2} to zero). In the balanced case, this model is equivalent to \textit{positive sequence}.
When modeling an asymmetric network and/or harmonics, the method can increase the minimum $\Delta t$, but its usefulness is limited since it results in a time-varying model. 

\ac{QSP} simulation software tools are highly developed for system-level balanced analysis within the positive sequence category. For instance, PTI PSS/E, GE PSLF, PowerTech Lab's TSAT, and PowerWorld, among others, are some of the tools that share common solution methods and models. The model libraries are generic and developed in coordination with \acp{ISO} standards and practices.

\vspace{-0.3cm}

\subsection{Electromagnetic Transients \label{sec:Sim_EMT}}
\ac{EMT} models are simulation models that can consider a diverse range of dynamic phenomena and are commonly implemented in studies that can significantly deviate from steady-state frequency. These are usually employed to study high-frequency phenomena, such as overvoltages, harmonic propagation, sub-synchronous resonance, and transient recovery voltages, among others \cite{martinez2014transient}.
Within this category, there is a large variety of modeling and computational approaches depending on the techniques used to limit the computational cost of executing the simulations.

\textit{1) Waveform simulation:}  A waveform simulation, also known as \textit{point-on-wave} simulation or $abc$-simulation, is the most comprehensive implementation of \ac{EMT} simulations. These simulation models represent the full wave throughout the entire simulation, which results in a time-variant model. As a result, these models require special consideration when initializing the simulation and choosing integration techniques. To capture fast electromagnetic phenomena, these simulations usually feature more detailed transmission line models than the $\pi$-model (as described in Section \ref{sec:Network}). A waveform simulation allows the use of distributed models such as the Bergeron or frequency-dependent models. A waveform simulation can include detailed converter switching dynamics; it uses similar methods but employs piece-wise continuous models between switching instants and requires additional equations and solution methods to capture converter transistors' ON/OFF switching behavior.

\textit{2) $\dqz$-model:} A $\dqz$-model is used to model the network in a  rotating reference frame commonly modeled using the \emph{transformed} $\pi$-model from \eqref{eq:dq_current}-\eqref{eq:dq_voltage2}. In the balanced case, also known as a \textit{fast time-varying phasor} simulation~\cite{venkatasubramanian1995fast}, the model results in a time-invariant formulation that yields the same initialization routine of the positive-sequence model, but the simulation can represent the network dynamics limited only by the $\pi$ model assumptions. Since the network is now modeled using only \acp{ODE} (i.e., there are no longer algebraic equations in the network model), the system model becomes a series of stiff \acp{ODE} because of the multi-rate nature of the system dynamics. As a result of the increased stiffness, finding the solution to the model might require the use of  Simultaneous-Implicit solution methods~\cite{stott1979power}. Most of the benefits of this formulation are lost when used in unbalanced networks or with harmonics since the model results in time-varying  signals that have similar requirements to \emph{waveform simulations}.

\textit{3) Dynamic Phasors:} This approach requires constructing that the model be formulated using Dynamic Phasors (see Section~\ref{sec:dynamic_phasor}). Dynamic phasors are useful when the frequency spectrum of power system transients are multiple narrow-banded signals centered around multiple harmonics of a fundamental frequency $\varrho$. The choice of $\varrho$ depends on the application; when employed to model converters, it could be the switching frequency, or in the case of system-level analysis, $\omega_s$ is a common choice. This approach can be implemented in several ways depending on the application, and if done correctly, the resulting model will be time-invariant but at the expense of additional states. A time-variant model enables other analytical techniques, such as sensitivity studies via small-signal analysis at equilibrium points making the approach practical beyond the formulation of simulation models. As shown in~\cite{demiray2008simulation}, dynamic phasors are helpful if the computational efficiency gained by reducing the model's bandwidth (and usage of adaptive time stepping techniques) outweighs the cost of increasing the number of states.

The most developed software solutions for \ac{EMT} simulations focus on waveform dynamics with and without converter switching and include models for very detailed component-level analysis. Examples of software environments are PSCAD, EMTP\textregistered, Simulink, PLECS, among others.

\subsection{Numerical integration methods considerations}

{In power systems applications, it is common to use numerical integration methods that, in part rely on the use of transformations and simplifications---thus embedding modeling assumptions into most simulation applications. In this subsection, we highlight important factors related to the numerical integration of simulation models as they relate to the model's formulation. See Chapter 2 in \hbox{\cite{cellier2006continuous}} for an in-depth look a choosing numerical integration methods for continuous-time simulation.
}

{The literature on numerical integration methods applied to specific simulation categories is extensive. \ac{QSP} simulations based on the index-1 \ac{DAE} model \eqref{eq:sim3_phasor}--\eqref{eq:sim4_phasor} can use the ``Partitioned-Solution method'' which solves the differential equations separately from the algebraic equations, or a ``Simultaneous-Solution method'' which jointly solves both differential and algebraic equations.  In~\hbox{\cite{stott1979power}}, B. Stott presents a detailed account of the solution techniques applicable to \hbox{\ac{QSP}} simulations and a historical account of why the field mostly converged to the use of partitioned methods.} {Waveform \ac{EMT} simulations mostly employ the numerical integration substitution technique with a trapezoidal rule for integration as originally developed by Dommel~\hbox{\cite{dommel1969digital}} to simulate circuit behavior. In~\hbox{\cite{watson2003power}}, the authors cover in detail the evolution of the ``Dommel Method" and how to address limitations like the modeling of nonlinear elements and switching.}

{Regardless of the simulation category, the model choices have impacts on the integration techniques. This is particularly critical when considering trade-offs between accuracy and solution speed.}

\subsubsection{Step size and integration approximation error}

{Consider the Taylor approximation of a trajectory for a model defined by \hbox{\eqref{eq:sim1}-\eqref{eq:sim2}} as follows (we simplified the notation by collecting the arguments of the function $F$ and $G$):}
\begin{subequations}
\begin{align}
    \vec{\boldsymbol{x}}(t + \Delta t) = \vec{\boldsymbol{x}}(t) +   F(\cdot, t)\Delta t + \frac{dF(\cdot, t)}{dt}\frac{\Delta t^2}{2!} + \dots \label{eq:taylor1}\\
   \vec{\boldsymbol{y}}(t + \Delta t) = \vec{\boldsymbol{y}}(t) +   G(\cdot, t)\Delta t + \frac{dG(\cdot, t)}{dt}\frac{\Delta t^2}{2!} + \dots \label{eq:taylor2}
\end{align}
\end{subequations}
\noindent {The approximation accuracy and computational cost of a numerical integration method depend on the step size, $\Delta t$, and the integration method order. Reducing $\Delta t$ can improve accuracy by reducing the relative importance of high-order terms, but it comes at the expense of more algorithm iterations. On the other hand, higher-order approximations tend to be more accurate and allow for a larger $\Delta t$ but require the evaluation of more terms. 
}
{As discussed in \cite{cellier2006continuous}, if the largest error tolerated within a single iteration is $10^{-n}$, it is best to choose at least a $n^\mathrm{th}$ order algorithm. One of the main motivations for using simplifications and transformation in simulation models is to enable larger $\Delta t$ without compromising the accuracy of the modeled states.}
{For example, reference-frame transformations and dynamic phasors allow the modeling of fast time-varying states using envelopes. This, in turn, permits larger time steps than those required by full waveform models.  Further, if the model complies with the requirements to employ \ac{SPT}, the differential terms that represent fast dynamics (e.g., \eqref{eq:spt_inverter1}-\eqref{eq:spt_inverter2}) can be eliminated, further increasing the required minimum $\Delta t$ required for a given accuracy, and accelerating computation.}

{Although simulation accuracy depends on the order and time step-length, most commercial software tools do not provide a mechanism to adjust the integrator order. As a result, time step reduction is the only option to adjust the simulation time and accuracy.}

\subsubsection{Model stiffness considerations}


{Although there is no widely accepted definition of model stiffness, a classical measure is the stiffness ratio, $\sigma$, defined as:}
\begin{equation}
    \sigma = \frac{\text{Re}(\lambda_\mathrm{max})}{\text{Re}(\lambda_\mathrm{min})}(\Delta t),
\end{equation}
{where $\text{Re}(\lambda_\mathrm{max})$ and $\text{Re}(\lambda_\mathrm{min})$ are the real parts of the largest and smallest eigenvalues, respectively. The challenges of simulating systems with \acp{IBR} stem from interactions across time scales, and the requirements to include fast electromagnetic dynamics increases model stiffness \cite{henriquez2020grid}.
Explicit integration methods, like Adams-Bashforth, widely used in \ac{QSP} simulation software \cite{concepcion2016extended}, have difficulty coping with highly stiff simulation models that result from the simultaneous modeling of electromechanical and electromagnetic phenomena. }

{Implicit numerical integration methods can overcome explicit methods' challenges with stiff systems. At each step of the simulation, an implicit algorithm solves a system of non-linear equations:}
\begin{align}
    S_F(\vec{\boldsymbol{x}}(t), \vec{\boldsymbol{y}}(t), \vec{\boldsymbol{x}}(t+ \Delta t), \vec{\boldsymbol{y}}(t+ \Delta t), \boldsymbol{\eta}) &= 0\\
    S_G(\vec{\boldsymbol{x}}(t), \vec{\boldsymbol{y}}(t), \vec{\boldsymbol{x}}(t+ \Delta t), \vec{\boldsymbol{y}}(t+ \Delta t), \boldsymbol{\psi}) &= 0.
\end{align}
{Newton's method is commonly used for the non-linear step in implicit methods; this reduces the process to a sequence of large linear system solves $Ax=b$ where $A$ is of size $n\times n$ ($n$ is the number of states). As a result, the computational complexity of an implicit method is dominated by the linear solution method, which is a function of the required precision and number of variables, with an upper bound $O(n^3)$. Several general-purpose integration algorithms use advanced stepping methods and parallel linear algebra libraries to reduce solution time in large stiff systems \hbox{\cite{bojanczyk1984complexity}}. The non-linear solve requirements of this formulation increase the computational burden per time step, but it enables solving systems that are not numerically stable with explicit methods.}

{Simulation transformations and simplifications can avoid the need for implicit solvers by generating a non-stiff representation of an otherwise stiff system. For example, if a model complies with the conditions to use \ac{SPT} it is possible to remove faster differential terms and reduce model stiffness. This technique has been used successfully in \ac{QSP} simulation software to eliminate stator electromagnetic dynamics from generator models \cite{sauer1988integral} as exemplified in Section IV.B. Reduced simulation models that can employ explicit methods and large step lengths will generally be several times faster than a stiff model solved with an implicit method.}

\subsubsection{Fixed and variable time step}

{The choice between fixed and variable time step solvers depends on the properties of the simulation model. Fixed-step integration methods use the same $\Delta t$ throughout a simulation and are the only suitable alternative for time-variant models. As discussed in the previous section, decreasing $\Delta t$ to increase accuracy also increases the simulation time.}
{However, the accuracy of fixed time step approaches depends on the system properties and it does not imply that the integration error remains constant throughout the simulation \cite{cellier2006continuous}. As a result, in fixed-time step solvers, the step length is usually decreased to meet the most stringent accuracy requisites, which might only be required during a portion of the simulation and results in unnecessary computations.}

{When a simulation model is time-invariant, it is possible to employ variable-step solution methods. Variable-step solvers employ different techniques and heuristics to change the step size throughout the simulation to reduce computing time without compromising accuracy. For example, during fast dynamics, variable-step solvers decrease the step size, and as the system settles and slower dynamics dominate, the step size increases. Although it is possible that the calculation of suitable step length throughout the simulation (i.e., step rejection) increases the computational cost, the additional cost is presumed to be offset by performing a reduced number of steps. The effectiveness of variable time-step methods depends on the implementation details making it difficult to develop generalized guidelines for their use. However, there have been algorithmic innovations in the variable time step control field to reduce rejection; notably, the use of \emph{PID} step-size control~\cite{gustafsson1994control}.}{In practice, most commercial applications employ fixed time step methods for both \ac{EMT} and \ac{QSP} simulations, and adaptive time stepping is seldom used. }

\subsubsection{Interfacing considerations}

\begin{figure}[t]
    \centering
    \includegraphics[width=0.45\textwidth]{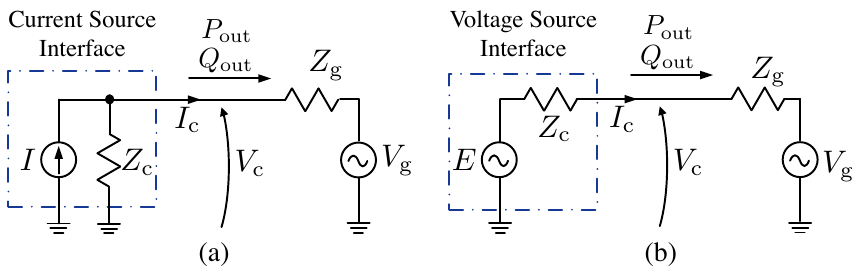}
    \caption{Interface options of devices to an infinite bus $V_g$.}
    \label{fig:interface}
\end{figure}

{Given the control flexibility encountered in \ac{IBR} control, these sources can be modeled as ``current source interface'' with a  large Norton impedance (see Fig.~\ref{fig:interface}(a)) or as a ``voltage source"  using a Thevenin equivalent (see Fig.~\ref{fig:interface}(b)). Although from the circuit perspective both models are equivalent, the interfacing modeling choice has implications for the numerical stability of the integration method.}

{When an \ac{IBR} is modeled as a current source, the assumption is that the Norton equivalent current is the same as the network-injected current. Current injection models can lead to non-convergence in the integration method due to small diagonal entries in the diagonal of the admittance matrix from the inverse of \acp{IBR} Norton impedance. When the admittance matrix is close to the singularity, it is more difficult to solve the network voltages (i.e., $\vec{\boldsymbol{i}} = \boldsymbol{Y} \vec{\boldsymbol{v}}$). Furthermore, the Norton injected current would be a function of the voltage at the terminals, so there needs to be a convergence routine at each time step when differential equations and algebraic equations are solved sequentially. Additionally, in cases with a weak grid, the convergence steps might not converge in a reasonable number of iterations. On the other hand, if both systems of equations are solved simultaneously, convergence-related issues are reduced, but in extreme cases, the Newton iteration can also fail if it results in a Jacobian bad numerical conditioning.}

{It is typical for device models in commercial computational tools to be connected to the network using a Norton current-source equivalent (see Fig.~\ref{fig:interface}). This is because inverter models are primarily designed for a current control representation. When the models have a ``voltage source interface,'' the inverter model is connected through a Thevenin impedance. However, this impedance is only included in the admittance matrix if the device model is changed to its Norton equivalent (see \cite{steps, siemens2017pss} for details on the conversion between Norton and Thevenin equivalents). To improve numerical convergence, using Thevenin interfaced models and a current control representation is more appropriate, but this requires carefully constructed IBR control models.}

\subsection{Exemplifying simulation cases}

\begin{figure}[t]
    \centering
    \includegraphics[width=0.45\textwidth]{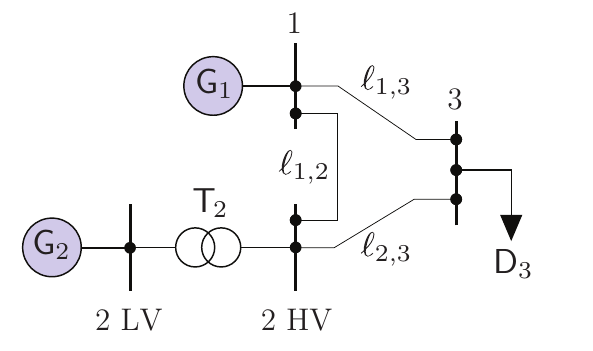}
    \caption{{Three-bus high-voltage (HV) network system motivating example.}}
    \label{fig:4bus_sys}
\end{figure}

{This section showcases the simulation results of a line trip between buses 1 and 2 for a small \ac{IBR} dominated system in Fig.~\hbox{\ref{fig:4bus_sys}} using three model formulations: waveform \ac{EMT}, $\dq$-EMT, and \ac{QSP}. The objective is to showcase the \hbox{$\dq$}--\hbox{\ac{EMT}} equivalence with waveform simulations and demonstrate the effects of \ac{QSP} simplifying assumptions. The network is modeled using a \hbox{$\pi$}-line, \hbox{$\mathsf{G}_1$} features a grid forming inverter with \ac{VSM} control \hbox{\cite{d2015virtual}} and \hbox{$\mathsf{G}_2$} uses droop-control \hbox{\cite{markovic2021understanding}}. The \ac{QSP} model formulation uses an algebraic representation of the electromagnetic circuits, lines, and inverter filters.}

{The \ac{QSP} and $\dq$--EMT simulations were conducted using the flexible formulation software \texttt{PowerSimulationsDynamics.jl} (\texttt{PSID.jl}), and the waveform simulation was conducted with PSCAD. The control blocks of the inverters are implemented as custom PSCAD components to match the models available in \texttt{PSID.jl} employing an ideal source output at the converter side of the filter.\footnote{The code to reproduce this experiment can be obtained at \url{https://github.com/Energy-MAC/SimulationModelsComparison.git}}} {From the numerical standpoint, the $\dq$-EMT model has a stiffness ratio of $\sigma = 2185$, whereas in the \ac{QSP} case, $\sigma = 105$. The order of magnitude difference between stiffness ratios and evaluation steps demonstrates the implications of including electromagnetics in the average model and the potential benefits of employing \ac{SPT} in the model formulation.}

The \hbox{$\dq$}--\hbox{\ac{EMT}} and \ac{QSP} models employ the high-precision adaptive time step solver Rodas5 \cite{rackauckas2017differentialequations} with $\text{abstol}=10^{-14}$ and $\text{reltol}=10^{-14}$. The waveform \ac{EMT} simulation is set to a 5$\mu$s fixed time-step. These values are used to guarantee that the comparisons are made at high accuracy levels. {To emphasize the computational impacts of the transformation and simplifications, over a 10-second simulation the fixed-time-step method performed $2\cdot 10^6$ evaluations, while the \hbox{$\dq$}--\hbox{\ac{EMT}} only required 31736 evaluations with time-steps in a range between 1.7$\mu$s and 7$m$s. However, only 18 of the the 31736 where smaller than 1$m$s. On the other hand, he simplest \ac{QSP} model only required 567 steps with time-steps in a range between 1$\mu$s to 0.1s making it the model with the least amount of evaluations required as expected given the discussion in Section IV.A}

\begin{figure}[t]
    \centering
    \includegraphics[width=0.9\columnwidth]{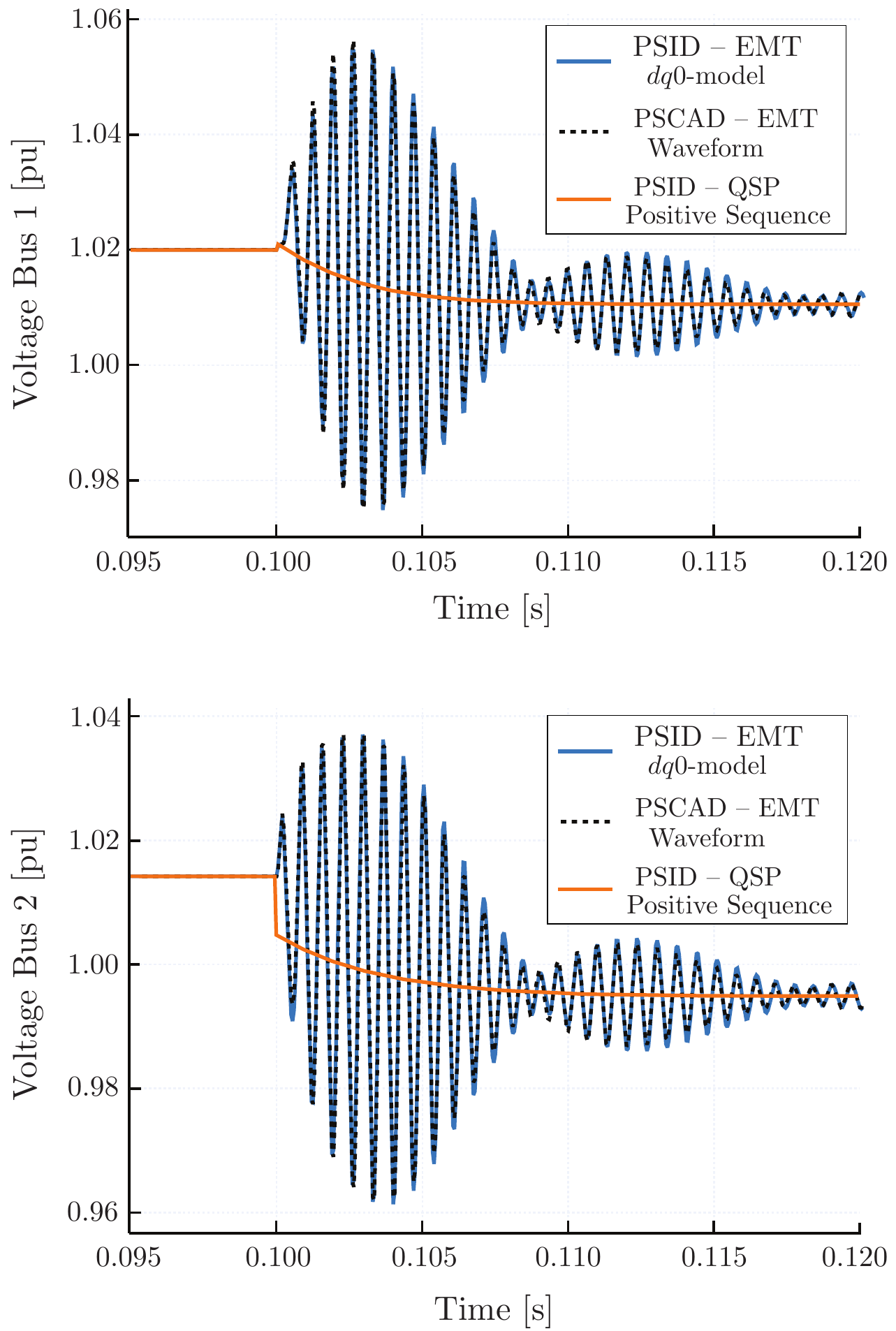}
    \caption{{Voltage magnitude transient response.}}
    \label{fig:pscad_voltages}
\end{figure}

\begin{figure*}[t]
    \centering
    \includegraphics[width=0.99\textwidth]{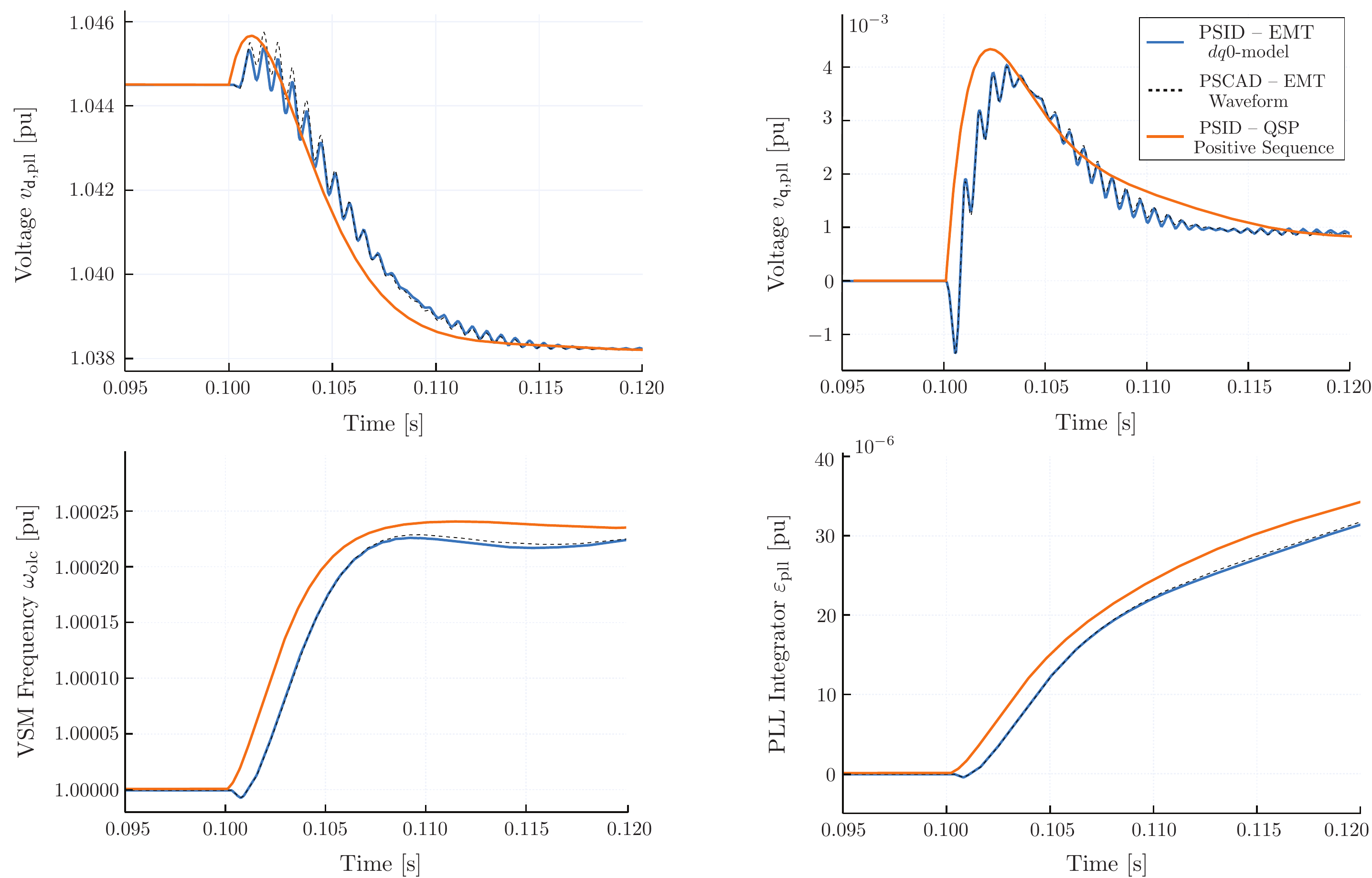}
    \caption{{Transient response for Inverter with \ac{VSM} control}}
    \label{fig:pscad_states}
\end{figure*}

{Figure \hbox{\ref{fig:pscad_voltages}} showcases the close match between $\dq$ and waveform solutions for the bus voltages and the capability of average models to capture \ac{EMT} dynamics. The results also show that the \hbox{\ac{QSP}} model does not capture high-frequency dynamics and has a slightly different trajectory. Figure \hbox{\ref{fig:pscad_states}} shows a comparison of the internal states of the \hbox{\ac{PLL}} and outer control in the inverter with \hbox{\ac{VSM}} control. These results display the effects of detailed electromagnetic modeling in the internal \hbox{\ac{PLL}} and \hbox{\ac{VSM}} states. The results showcase that the additional dynamics from the lines and the filter's electromagnetics affect the \hbox{\ac{VSM}} frequency estimation $\omega_\text{olc}$, and in extreme cases, the oscillations could destabilize the controller. Further, both \hbox{\ac{EMT}} models can reflect the oscillatory dynamics of the states $v_{\d, \text{pll}}$ and $v_{\q, \text{pll}}$, which demonstrates the equivalence between waveform and $\dq$ models {for three-phase balanced systems}.}

\section{Concluding Remarks} \label{sec:Conclusions}

We have made the case that the distinction between \ac{QSP} and \ac{EMT} models is insufficient to characterize the fundamentals of specific simulation models and techniques. Instead, as we reviewed in this paper, there is a large diversity of transformations and simplifications that yield a range of simulations models, each of which differently balances computational complexity with model fidelity---provided that certain operating conditions are met, such as narrow-band phenomena in dynamic phasors. Compared with a computationally expensive waveform simulation, these transformations enable cheaper \ac{EMT} simulations without loss of information. Figure~\ref{fig:classification} and Table \ref{tab:category} summarize the proposed classification taxonomy for time-domain simulations.

This paper also showcases that the applicability of certain simplification techniques, like \ac{SPT}, relies heavily on the control parameters of \acp{IBR} and the time constants of the remainder of the system. As a result, modeling requirements must be determined from a detailed model before any simplifications. This, in turn, affects the generalizability of specific device models and assumptions in different systems under different disturbances.

{When selecting a simulation model and numerical method, the question to answer is: ``What constitutes a valid solution?" In some situations, the oscillatory behavior observed in Fig.~\ref{fig:pscad_voltages} is not relevant to the analysis, and it might not be important to recover that information. This would be especially true over long simulation time spans. However, in other situations, information about high-frequency oscillation (such as its magnitude and energy) is highly relevant. This indicates a careful selection of which states to maintain and which ones to eliminate is critical. Finally, there will be situations like protection switching studies or transformer energizing where the waveform results will be required, and average models might not suffice to recover critical, very fast transient behavior.}

Finally, we stress that for practitioners, the distinction between simulation models comes from the availability of a framework to solve such simulations. The lack of commercial software for novel simulation techniques hinders the exploration of novel alternatives that address existing approaches' shortcomings. {Thus, research and development on time-domain simulation software are essential to analyzing, in a computationally efficient manner, the {myriad \hbox{\acp{IBR}} architectures} of future AC-DC hybrid power systems. }


\section*{Acknowledgements}

We thank C. Roberts, M. Bossart, G. Colon-Reyes, B.M. Hodge, and many others for fruitful discussions that led to the development of this paper. We thank M. Bossart for helping implement the PSCAD simulations used in the example. This work was authored in part by the National Renewable Energy Laboratory, operated by Alliance for Sustainable Energy, LLC, for the U.S. Department of Energy (DOE) under Contract No. DE-AC36-08GO28308. Funding provided by U.S. Department of Energy's Office of Energy Efficiency and Renewable Energy under Solar Energy Technologies Office Agreement Number 33505.
The views expressed in the article do not necessarily represent the views of the DOE or the U.S. Government. The U.S. Government retains and the publisher, by accepting the article for publication, acknowledges that the U.S. Government retains a nonexclusive, paid-up, irrevocable, worldwide license to publish or reproduce the published form of this work, or allow others to do so, for U.S. Government purposes.

\bibliographystyle{IEEEtran}
\bibliography{main_v7}

\begin{IEEEbiographynophoto}{José Daniel Lara (Senior Member, IEEE)}
received the B.Sc. and Licentiate degrees in electrical engineering from the University of Costa Rica, San Pedro, Costa Rica, in 2009 and 2012, respectively, the M.Sc. degree in Electrical and Computer Engineering from the University of Waterloo, Waterloo, ON, Canada, in 2014,  M.S. in Energy and Resources with focus on Engineering and Business for Sustainability in 2017, and Ph.D. in Energy and Resources with Designated Emphasis in Computational and Data Science and Engineering from the University of California, Berkeley, CA, USA in 2022. He currently works at the National Renewable Energy Laboratory (NREL) with the Grid Planning and Analysis Center (GPAC). His research interests include the analysis and simulation of power systems and their applications to power systems operation and quantitative energy policy.
\end{IEEEbiographynophoto}

\begin{IEEEbiographynophoto}{Rodrigo Henriquez-Auba (Member, IEEE)}
received the B.Sc. and M.Sc. in electrical engineering from the Pontifical Catholic University of Chile, Santiago, Chile, in 2014 and 2016, respectively, and the Ph.D. degree in Electrical Engineering and Computer Sciences from the University of California, Berkeley, CA, USA, in 2022. He currently works at the National Renewable Energy Laboratory (NREL) and his research interests include operation and planning of electric systems with large shares of renewable sources, and control, modeling and simulation of power systems and power electronics.
\end{IEEEbiographynophoto}

\begin{IEEEbiographynophoto}{Deepak Ramasubramanian (Senior Member, IEEE)}
received the B.E. degree from the PES Institute of Technology, Bangalore, KA, India, in 2011, and the M.Tech. degree from the Indian Institute of Technology Delhi, New Delhi, India, in 2013 and Ph.D. degree at Arizona State University, Tempe, AZ, USA in 2017. Deepak is a Technical Leader in the Grid Operations and Planning Group at EPRI and leads research projects related to modeling of inverter-based resources for bulk power system analysis.
\end{IEEEbiographynophoto}
\vskip -2\baselineskip plus -1fil
\begin{IEEEbiographynophoto}{Sairaj Dhople (Senior Member, IEEE)} received the B.S., M.S., and Ph.D. degrees in electrical engineering from the University of Illinois at Urbana-Champaign, Urbana, IL, USA, in 2007, 2009, and 2012, respectively. He is currently Robert \& Sydney Anderson Associate Professor with the Department of Electrical and Computer Engineering, University of Minnesota, Minneapolis, MN, USA. His research interests include modeling, analysis, and control of power  electronics and power systems with a focus on renewable integration. Dr. Dhople is the recipient of the National Science Foundation CAREER Award in 2015, the Outstanding Young Engineer Award from the IEEE Power and Energy  society in 2019, and the IEEE Power and Energy Society Prize Paper Award in 2021.
\end{IEEEbiographynophoto}
\vskip -2\baselineskip plus -1fil
\begin{IEEEbiographynophoto}{Duncan S. Callaway (Member, IEEE)}
is an Associate Professor of Energy and Resources at the University of California, Berkeley.  He is also a faculty affiliate in Electrical Engineering and Computer Science, and a faculty scientist at Lawrence Berkeley Laboratory.  He received his PhD from Cornell University.  He has held engineering positions at Davis Energy Group and PowerLight Corporation, and academic positions at UC Davis, the University of Michigan and UC Berkeley.  Duncan teaches courses on electric power systems and at the intersection of statistical learning and energy. His research focuses on grid integration of renewable electricity, and models and control strategies for power system dynamics, demand response, electric vehicles and electricity storage.
\end{IEEEbiographynophoto}
\vskip -2\baselineskip plus -1fil
\begin{IEEEbiographynophoto}{Seth Sanders (Fellow, IEEE)}
received the S.B. degree in electrical engineering and physics in 1985, and the S.M. and Ph.D. degrees in electrical engineering from the Massachusetts Institute of Technology, Cambridge, MA, USA, in 1985 and 1989, respectively.
He is currently a Professor with the Department of Electrical Engineering and Computer Sciences, University of California, Berkeley, CA, USA, and co-founder and Chief Technology Officer with Amber Kinetics, a Technology Developer and manufacturer
of utility scale flywheel energy storage systems. Following an early experience as a Design Engineer with the Honeywell Test Instruments Division from 1981 to 1983, he joined the UC Berkeley faculty in 1989. His technical interests are broadly in electrical energy and power conversion systems. He is presently or has recently been active in supervising research projects in the areas of renewable energy systems, high frequency integrated power conversion circuits, IC designs for power conversion applications, and electric machine systems. During the 1992–1993 academic year, he was on industrial leave with National Semiconductor, Santa Clara, CA.
Dr. Sanders was the recipient of the NSF Young Investigator Award and multiple Best Paper Awards from the IEEE Power Electronics and IEEE Industry Applications Societies, and IEEE PELS Modeling and Control Technical Achievement Award. He was the Chair of the IEEE PELS Technical Committee on Computers in Power Electronics, Chair of the IEEE PELS Technical Commit- tee on Power Conversion Components and Systems, and as Member-At-Large of the IEEE PELS Adcom. He is a past Distinguished Lecturer of the IEEE PELS and IAS societies.
\end{IEEEbiographynophoto}

\end{document}